\documentclass[acmsmall]{acmart}
\pdfoutput=1

\usepackage[ruled,linesnumbered]{algorithm2e}
\usepackage{amsmath}
\usepackage{booktabs} % For formal tables
\usepackage{color}
\usepackage{multirow}
\usepackage{graphicx}
\usepackage{subfigure}
\usepackage{balance}
\usepackage{multicol}
\usepackage{bbding}
\usepackage{makecell}
\usepackage{amstext}
\usepackage{setspace}
\usepackage[figuresright]{rotating}
\usepackage{lscape}
\usepackage{adjustbox}
\usepackage{capt-of}
\usepackage{enumerate}
\usepackage{bm}

\AtBeginDocument{%
  \providecommand\BibTeX{{%
    \normalfont B\kern-0.5em{\scshape i\kern-0.25em b}\kern-0.8em\TeX}}}

\setcopyright{acmcopyright}
\acmJournal{TOIS}
\acmYear{2021} \acmVolume{1} \acmNumber{1} \acmArticle{1} \acmMonth{1} \acmPrice{15.00}\acmDOI{10.1145/3466753}

\begin{document}

\title{Learning to Learn a Cold-start Sequential Recommender}

\author{Xiaowen Huang}
\email{xwhuang@bjtu.edu.cn}

\author{Jitao Sang}
\email{jtsang@bjtu.edu.cn}

\author{Jian Yu}
\email{jianyu@bjtu.edu.cn}

\affiliation{%
  \institution{School of Computer and Information Technology \& Beijing Key Lab of Traffic Data Analysis and Mining, Beijing Jiaotong University}
%  \streetaddress{95 Zhongguancun Rd}
  \city{Haidian Qu}
  \state{Beijing Shi}
  \country{China}}

\author{Changsheng Xu$^{*}$}\thanks{Corresponding author: Changsheng Xu}
\email{csxu@nlpr.ia.ac.cn}
\affiliation{%
  \institution{National Lab of Pattern Recognition, Institute of Automation, Chinese Academy of Sciences}
  \streetaddress{95 Zhongguancun Rd}
  \city{Haidian Qu}
  \state{Beijing Shi}
  \country{China}}

\affiliation{%
  \institution{School of Artificial Intelligence, University of Chinese Academy of Sciences}
  \streetaddress{80 Zhongguancun Rd}
  \city{Haidian Qu}
  \state{Beijing Shi}
  \country{China}}
  
\affiliation{%
  \institution{Peng Cheng Laboratory}
  \city{Shenzhen}
  \country{China}} 

\begin{abstract}

The cold-start recommendation is an urgent problem in contemporary online applications. It aims to provide users whose behaviors are literally sparse with as accurate recommendations as possible. Many data-driven algorithms, such as the widely used matrix factorization, underperform because of data sparseness. This work adopts the idea of meta-learning to solve the user's cold-start recommendation problem. 
We propose a meta-learning based cold-start sequential recommendation framework called metaCSR, including three main components: Diffusion Representer for learning better user/item embedding through information diffusion on the interaction graph; Sequential Recommender for capturing temporal dependencies of behavior sequences; Meta Learner for extracting and propagating transferable knowledge of prior users and learning a good initialization for new users. metaCSR holds the ability to learn the common patterns from regular users' behaviors and optimize the initialization so that the model can quickly adapt to new users after one or a few gradient updates to achieve optimal performance. The extensive quantitative experiments on three widely-used datasets show the remarkable performance of metaCSR in dealing with user cold-start problem. Meanwhile, a series of qualitative analysis demonstrates that the proposed metaCSR has good generalization.
 
\end{abstract}




\maketitle

Recommendation systems (RS) intend to address the information explosion by finding a set of items for users to meet their personalized interests in many online applications, such as E-commerce websites \citep{linden2003amazon}, social networks \citep{kywe2012a}, video-sharing sites \citep{covington2016deep} and news websites \citep{wang2018dkn}. In the decades since the rapid development of RS, many effective recommendation algorithms have been proposed: from the content-based methods, to the widely used collaborative filtering (CF) algorithm, to the recently emerging deep learning-based approaches. Although these kinds of recommendation algorithms generally work well when sufficient data is available, cold-start recommendation which address the sparseness problem is yet a difficult and urgent problem to be solved in practical applications. 

Cold-start happens when new users/items arrive on online platforms. Classic recommendation methods like CF assumes that there are sufficient user-item interactions (browse, click, rate, etc.) for matrix factorization. However, new users/items usually contain extremely sparse data, 
% so that the typical matrix factorization techniques are inapplicable to achieve the desirable recommendation results.
so that we can infer ratings of similar users/items even if those ratings are unavailable. However, for new users/items, this becomes difficult because we have no or just only a few such interactive data for them. As a result, we cannot “fill in the blank” using typical matrix factorization techniques to achieve the desirable recommendation results due to the data sparseness issues.
Content-based methods \citep{lops2011content} usually rely on additional side information, for instance, user-specific demographics features (e.g., gender, location, nationality, religion) or item-specific properties (e.g., genre, publication year, actors, director in the case of movies). But in real-world scenarios, such information is usually missing due to the unavailability of data or user-privacy issues, which greatly reduces the effect of these methods.
Some previous efforts \citep{man2017cross, hu2018conet} designed the cross-domain recommender to tackle the cold-start problem with domain knowledge transfer. But they still require a large amount of shared samples across domains. 
Data sparseness issues, and the accessibility of additional side information remain key barriers to cold-start recommendations. 

This work addresses the cold-start issues in the sequential recommendation (SR) scenario. Compared with the traditional recommender, SR system holds the ability of capturing the evolution of users' dynamic interests \citep{huang2018csan, huang2019explainable}. In addition, there are some common patterns in the users' sequential behaviors. For example, when watching a movie or TV series, users usually watch the first episode before watching the sequels; when shopping on the E-commerce websites, users usually buy a new computer, and then a mouse, keyboard, and other accessories. The sequential patterns are common to almost all users. It is very helpful to assist recommendation by transferring the prior common patterns to cold-start users. 

We notice that the user cold-start recommendation problem can be formulated as a few-shot learning problem, where the meta-learning method is a recognized solution. A popular meta-learning method, Model-Agnostic Meta-Learning (MAML), provides a promising way to extract and propagate transferable knowledge of prior tasks and learn a good initialization for new tasks. In recent years, some work has introduced meta-learning algorithms into cold-start recommendation, but most of these algorithms need additional side information, and they do not model the temporal relationship of user behaviors, resulting in lacking the ability to model behavioral sequential patterns and the ability to capture user dynamic preferences.

To consider the characteristics of sequential recommender and address the limitations of existing cold-start recommending methods, we propose the \textbf{meta}-learning based \textbf{C}old-start \textbf{S}equential \textbf{R}ecommendation framework (metaCSR), an end-to-end framework for user cold-start sequential recommendation (CSR), which takes the user-item interactions from warm-start (regular) users with adequate behaviors as input, and outputs ``next-one'' item prediction for cold-start (new) users with few behaviors.
The key insight behind metaCSR framework is to learn the common patterns from regular users' behaviors, facilitate the initialization of cold-start users so that the model can quickly adapt to new users after one or a few gradient updates to achieve optimal performance.

There are three key components of metaCSR:
(1) The \textit{Diffusion Representer}, which works on the user-item interaction graph, is proposed to learn the users' and items' high-order interactive representation. The purpose of this module is to learn more effective user/item embedding only through the information diffusion on user-item interaction graph without using additional side information. 
(2) The \textit{Sequential Recommender}, which is based on self-attention mechanism, is used to model the temporal dependencies of users' sequential behaviors to capture users' dynamic interests.
(3) The \textit{Meta Learner}, which is a model-agnostic meta-learning algorithm, is employed to learn common patterns of user behaviors so that it can quickly adapt to new users with only a few gradient updates. 

% Improving cold-start sequential recommendation performance is conducive to improve the user experience, promote the user retention rate.

The contributions of this work are summarized as follows:
\begin{itemize}
	\item We propose a novel meta-learning based cold-start sequential recommendation framework, which is an end-to-end framework that can extract and propagate transferable knowledge of regular users and quickly adapt to new users.

	\item The incorporation of Diffusion Representer and Sequential Recommender helps to better capture users' dynamic preferences and achieve promising performance in dealing with user CSR problem without relying on any additional side information.

	\item We run extensive experiments on three real-world datasets. The promising results demonstrate the efficacy of our proposed metaCSR in addressing user CSR problem, while maintaining competitive performance in both warm-start and cold-start recommendation scenarios.
\end{itemize}

\section{Related Work}
In this section, we briefly review the related work that are most relevant to our work, including meta-learning methods, sequential recommendation methods and the existing recommendation approaches for the cold-start problem.

\subsection{Meta-learning}

\noindent Meta-learning, also called ``learning to learn'', is a recently popularized paradigm for training an easily generalizable model that can rapidly adapt to new tasks from only a few examples \citep{vilalta2002a}. There are three main research directions of meta-learning, including metric-based, model-based and optimization-based meta-learning. Metric-based meta-learning, such as matching network \citep{vinyals2016matching, snell2017prototypical}, aims to learn the similarity between samples within tasks. Model-based meta-learning model updates its parameters rapidly with a few training steps, which can be achieved by its internal architecture or controlled by another meta-learner model. Memory-augmented neural networks \citep{ravi2017optimization} and meta networks \citep{munkhdalai2017meta} are the typical model-based meta-learning methods. Optimization-based meta-learning intends for adjusting the optimization algorithm so that the model can be good at learning with a few examples, including MAML \citep{finn2017model}, Meta-SGD \citep{li2017meta} and Reptile \citep{nichol2018reptile:}, etc. These methods promise to extract and propagate transferable representations of prior tasks. 

\subsection{Sequential Recommendation}

Sequential recommendation (SR) problem is usually cast as sequence prediction problem. The sequence modeling methods mainly belong to Markov Chain based models \citep{rendle2010factorizing,cheng2013you} and RNN-based models \citep{hidasi2015session,hidasi2016parallel,quadrana2017personalizing,yu2016dynamic}.
Inspired by the great power of Matrix Factorization (MF), Factorized Personalized Markov Chain (FPMC) combines the power of MF and MC to factorize the transition matrix over underlying MC to model personalized sequential behaviors for the problem of next-item recommendation given the last-N interactions of the user \citep{rendle2010factorizing,cheng2013you}.
RNN computes the current hidden state from the current input in the sequence and the hidden state outputted by the previous time step. The recurrent feedback mechanism memorizes the influence of each past data sample in the hidden state. It therefore makes RNN and its variants such as LSTM and GRU be able to model the temporal information for user behaviors in recommendation task \citep{hidasi2015session,hidasi2016parallel,quadrana2017personalizing,yu2016dynamic}. Though it is an effective way to encode users' behavior sequences, it still suffers from several difficulties, such as hard to parallelize, time-consuming, hard to preserve long-term dependencies. The emergence of the Transformer \citep{vaswani2017attention} architecture tackles the problem of sequence transduction. Some studies abandon the complex and time-consuming RNN structures, and instead construct the sequence model based on self-attention mechanism and apply it to SR system. ATRank \citep{zhou2018atrank} takes the lead in using self-attention structure for the SR and achieves encouraging results. CSAN \citep{huang2018csan} adopts a feature-wise masked self-attention to construct user preference representations for SR. EIUM \citep{huang2019explainable} introducing knowledge graph into self-attention based Sequential Recommender models for providing explainable recommendations.

\begin{table}
  \caption{Main notations.}
  \label{tab:notation}
\begin{tabular}{l|l}
\hline\hline
{\bf Symbol} & {\bf Notation}\\
\hline\hline
$\mathcal{U}$ & user set \\
\hline
$\mathcal{U}_{reg.}$ & regular user set\\
\hline
$\mathcal{U}_{new}$ & new user set \\
\hline
$\mathcal{I}$ & item set\\
\hline
$ u  $ & a specific user\\
\hline
$ i $ & a specific item\\
\hline
$\mathcal{B}_u$ & user $u$'s historical behavioral sequence\\
\hline
$\mathcal{G=(V,E)}$ & user-item interaction graph\\
\hline
$\mathcal{V}$ & the vertex set\\
\hline
$\mathcal{E}$ & the edge set\\
\hline
$f_v$ & entity $v$'s representation\\
\hline
$f^{in}$ & inherent feature\\
\hline
$f^{la}$ & latent feature\\
\hline
$cor.(i_m,i_n)$ & correlation between $i_m$ and $i_n$\\
\hline
$M_{m,n}^{fw}, M_{m,n}^{bw}$ & position encoding matrics\\
\hline
$att.(i_m,i_n)$ & attention score between $i_m$ and $i_n$\\
\hline
$\textbf{s}_u $ & user $u$'s preference embedding\\
\hline
$ \mathcal{L} $ & loss function \\
%\hline
%$  $ &  \\
%\hline
%$  $ &  \\
%\hline
%$  $ &  \\
\hline \hline
\end{tabular}
\end{table}

\subsection{Cold-start Recommendation}

\noindent The research of cold-start recommendation mainly focuses on two aspects, named user cold-start recommendation \citep{pandey2016resolving} and item cold-start recommendation \citep{vartak2017a, houlsby2014cold, zhu2020addressing, pan2019warm}, which recommends for new users who have no/few historical behaviors, or recommends new products that have just been added to the system to the suitable users.
These methods suffer from a common issue that they need user demographics or item attributes to assist in modeling to be applied to cold-start recommendation tasks. 
However, the missing of additional side information due to unavailability of data or user privacy issues will greatly reduces the effect of these methods that rely on attribute information. 
Actually, the cold-start recommendation task is a natural few-shot learning problem. Since meta-learning is a powerful way to solve the few-shot learning problem, in recent years, some research works have introduced the idea of meta-learning into the cold-start recommendation task. \citet{vartak2017a} presents a meta-learning strategy to address item cold-start when new items arrive continuously. 
\citet{bharadhwaj2019meta} and \citet{lee2019melu} both designed a cold-start user recommendation model based on the MAML algorithm to rapidly adopt new users with a few examples, which is named MetaCS and MeLU, respectively. \citet{wei2020fast} presented a metaCF method which applicable to any differentiable CF-based models like FISM \citep{kabbur2013fism} and NGCF \citep{wang2019neural}, to learn a suitable model for initializing the adaption. SML \citep{zhang2020retrain} is a sequential meta-learning method which offers a general training paradigm, where a neural network-based transfer component can transform the old model to a new model that is tailored for future recommendations.
\citet{du2019sequential} combine the scenario-specific learning with meta-learning for online-recommendation. \citet{zhao2019learning} develop a two-stage meta-learning algorithm to learn fixed and adaptive parts of model parameters for user cold-start recommendation. \citet{luo2020metaselector} propose a meta-learning framework to facilitate user-level adaptive model selection in recommendation system. 

The major difference between metaCSR and existing literature is that: we focus on the cold-start sequential recommendation task where common patterns of sequential behaviors are mined and learning through our meta-learning based algorithm. Moreover, Our proposed metaCSR is a general framework for CSR, which does not require any additional side information other than user ID, item ID, and interaction matrix of users on items, and can still achieve good results on the CSR task.

\section{PRELIMINARIES}
\subsection{Problem Formulation}

General user behaviors can be interpreted using the binary relationship between a user and an item.
We denote $\mathcal{U}=\{u_1,u_2,...,u_{|\mathcal{U}|}\}$ as the set of users, including regular user set $\mathcal{U}_{reg.}$ and new user set $\mathcal{U}_{new}$. $\mathcal{U} = \{\mathcal{U}_{reg.} \cup \mathcal{U}_{new} \ \vert \ \mathcal{U}_{reg.} \cap \mathcal{U}_{new} = \emptyset \}$.  $\mathcal{I}=\{i_1,i_2,...,i_{|\mathcal{I}|}\}$ as the set of items.
Hence, the historical sequential records of user $u$ can be represented as $\mathcal{B}_u=\{ i_{t},t = 1,2,...,T\}$.
Based on these preliminaries, we are ready to define the sequential recommendation task: given $\mathcal{B}_u=\{i_{t},t = 1,2,...,T\}$ of a user towards items, the task is to predict the next item $i_{T+1}$ that the user may interact with at time $T+1$.

\newtheorem{myDef}{Scenario}
\newtheorem{myTheo}{Theorem}
\begin{myDef}
Cold-start scenario.

Our main purpose is to propose a universal framework to address user cold-start recommendation problem, so the main task of this work is to train the model on $\mathcal{B}_{u \in \mathcal{U}_{reg.}}$, and perform sequential recommendation on new users $u \in \mathcal{U}_{new}$, which is called user cold-start scenario.
\end{myDef}

\begin{myDef}
Warm-start scenario.

We also want to know whether it is effective in dealing with general recommendation task, thus we also conduct the experiments in user warm-start scenario, where we still make the predictions on regular users $u \in \mathcal{U}_{reg.}$.
\end{myDef}

% Our main purpose is to propose a universal framework to address user cold-start recommendation problem, so the main task of this work is to train the model on $\mathcal{B}_{u \in \mathcal{U}_{reg.}}$, and perform sequential recommendation on new users $u \in \mathcal{U}_{new}$, which is called user \textbf{cold-start scenario}. Besides, we also want to know whether it is effective in dealing with general recommendation task, thus we also conduct the experiments in user \textbf{warm-start scenario}, where we still make the predictions on regular users $u \in \mathcal{U}_{reg.}$.

\subsection{Notations}
The main notations of this work are summarized in Table~\ref{tab:notation}.
%We have omitted some temporary symbols that only appear once.

\begin{figure*}[!htbp]
    \centering
    \includegraphics[width=1\textwidth]{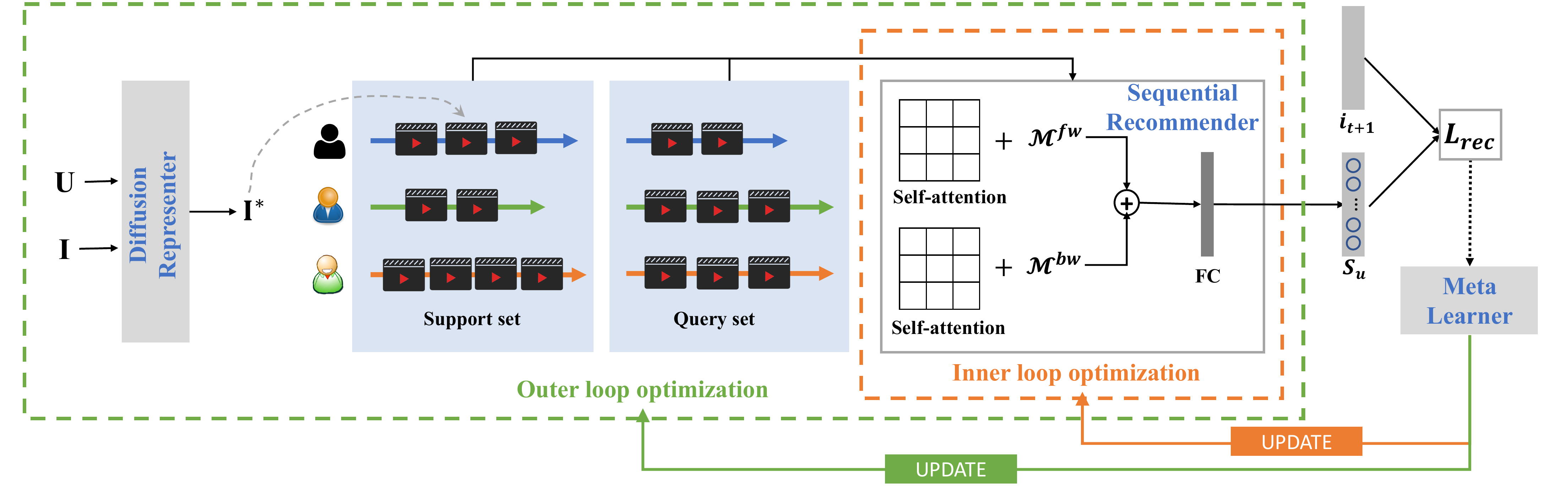}
    \caption{Illustration of meta-learing based cold-start sequential recommendation framework (metaCSR).}
    \label{fig:framework}
\end{figure*}

\section{Methodology}

In this section, we present the technical details of \textbf{metaCSR} algorithm for cold-start sequential recommendation.
The overall framework is illustrated in Fig. ~\ref{fig:framework}, where three key components are involved in, including \textit{Diffusion Representer}, \textit{Sequential Recommender} and \textit{Meta Learner}. 

\subsection{Diffusion Representer}

There are many ways for the embedded representation of users and items, such as the simplest one-hot representation, or low-dimensional dense vectors which are compressed from high-dimensional sparse vectors through an embedding layer (such as a fully connected network). Moreover, fusion representations by introducing auxiliary information, such as user's attributes, item's text and images, etc. Furthermore, there are semantic representations that contain semantic graph structure information, for example, introducing knowledge graphs into the representation learning of users and items. No matter what kind of representation manners, the ultimate goal is to represent users and items better.

In recommendation tasks, mining users' interest is the core means to improve task performance. The development of learning on graph-structured data, which is fundamental for recommendation applications, for example, to exploit user-item interaction graphs as well as social graphs~\citep{ying2018graph}. Several recommendation models utilize user’s local neighbors’ preferences to alleviate the data sparsity issue. Graph Convolutional Networks (GCNs) have shown promising results by modeling the information diffusion process in graphs that leverage both graph structure and node feature information. Thus, we construct a user-item interaction graph $\mathcal{G}$ to modal the information diffusion effect in a more fine-grained way. We learn the users' and items' high-order interactive representations through aggregating neighbors' information, and apply them to the downstream tasks.

\begin{figure*}[!htbp]
    \centering
    \includegraphics[width=1\textwidth]{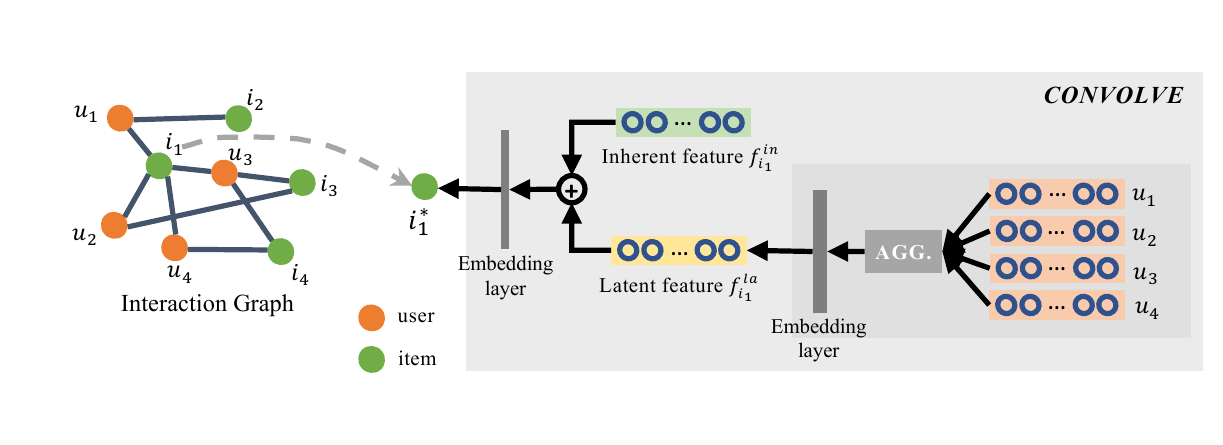}
    \caption{Illustration of Diffusion Representer.}
    \label{fig:gcn}
\end{figure*}

As illustrated in Fig. \ref{fig:gcn}, the user-item interaction graph $\mathcal{G}=\mathcal{<V,E>}$ is a bipartite graph composed of users $u\in \mathcal{U}_{reg.}$ and items $i\in\mathcal{I}$, the vertex/entity set $V = \mathcal{U}_{reg.} \cup I$, where users and items are connected through an interactive relationship.  
The Diffusion Representer, which works on the user-item interaction graph, is to learn the users’ and items’ high-order interactive representation. The purpose of this module is to learn more effective user/item embedding only through the information diffusion on user-item interaction graph without using additional side information. 
Certainly, additional auxiliary information (eg., textual and visual features of users and items) also can be seamlessly integrated into the graph.

Inspired by SocialGCN \citep{wu2018socialgcn}, the entity $v$'s representation $f_v$ in the graph is calculated from two parts, one is the \textit{inherent feature} vector of the entity itself, which can incorporate one-hot ID, attributes, context information, and so on. Let $f^{in}$ denotes the inherent feature. The other is the \textit{latent feature} vector obtained through information diffusion, which contains the structural information of the interaction graph and the information supplement of neighbor nodes, which is represented by $f^{la}$.

For each entity $v\in\mathcal{V}$, given its N-hop relevant entities, i.e. neighbor nodes $\mathcal{N}_v$, $v$ can be defined as:
\begin{equation}
    f_v = CONVOLVE(f^{in},f_{\hat{v}}, \hat{v}\in\mathcal{N}_v)
\end{equation}

To simplify the description, we take 1-order convolution as an example. The procedure is detailed in Algorithm \ref{alg:convolve}. The basic idea is that:
\begin{enumerate}[\hspace{1em} step 1.]
    \item An aggregator $AGG.$ (e.g., mean pooling, max pooling, etc.) is applied on the neighbors $\hat{v}\in\mathcal{N}_v$ to aggregate the influences from neighbors' representations.
    \item The aggregated representation $h_{\hat{v}}$ is transformed through an embedding layer to generate the latent feature embedding $f_v^{la}$ of entity $v$, the purpose of which is to project the aggregated vector $\hat{v}$ into the same space as the inherent feature embedding $f_v^{in}$.  The inherent feature $f_v^{in}$ is initialized via lookup operation. 
    \item The aggregated representations $h_{\hat{v}}$ through an embedding layer to generate the latent feature embedding $f_v^{la}$ of entity $v$, the purpose of which is to project the aggregated vector $\hat{v}$ into the same space as the inherent feature embedding $f_v^{in}$. 
    \item The normalization makes the training more stable. 
\end{enumerate}
A single convolution operation transforms and aggregates feature information from an entity's one-hop graph neighborhood, and by stacking multiple such convolutions information can be propagated across far reaches of a graph.

\begin{algorithm}
    \caption{CONVOLVE}
    \label{alg:convolve}
    \KwIn{Inherent feature embedding $f^{in}_v$ of entity $v$; A set of neighbor representation \{$f_{\hat{v}} \vert \hat{v}\in\mathcal{N}_v$\}}
    \KwOut{New embedding $f^*_v$ for entity $v$}
    
    $h_{\hat{v}} \gets AGG.( f_{\hat{v}} , \hat{v}\in\mathcal{N}_v)$;
    
    $f^{la}_v \gets ReLU(W_1h_{\hat{v}}+b_1)$;
    
    $f_v \gets ReLU(W_2 \cdot CONCAT(f^{in}_v,f^{la}_v)+b_2)$;
    
    $f^*_v \gets f_v/\Vert f_v \|_2$

    %\end{algorithmic}
\end{algorithm}

\subsection{Sequential Recommender}

Modeling users' sequential behaviors, capturing the dependencies between the elements of behavior sequence, is conducive to learn users' dynamic interests, thereby improving the performance of the sequential recommendation system. In order to simplify the structure of the model and improve the efficiency of training, inspired by ~\cite{vaswani2017attention,zhou2018atrank,huang2018csan,huang2019explainable}, we adopt the masked self-attention network, an element-wise self-attention module with two position encoding matrices $M_{m,n}^{fw}$ and $M_{m,n}^{bw}$, to model the sequential behaviors, in which we can benefit not only from the ability in capturing the long-distance dependencies of the sequence with various lengths, but also from the capability in parallel computing for efficiently learning.

For each user $u$'s historical behavior sequence $\mathcal{B}_u = \{i_{t},t = 1,2,...,T\}$, where $T$ is the length of the sequence, element $i_t$ is represented by item feature $f_i$ which is obtained from Diffusion Representer module. The masked self-attention network takes $\mathcal{B}_u$ as input, and then encodes the sequence to obtain the final semantic representations of the entire sequence. They will be updated iteratively with the training of the model. Take the sequence $\mathcal{B}_u$ as input, $cor.(i_m, i_n)$ denotes the correlations between the two elements:
\begin{equation}
cor.^{fw/bw}(i_m,i_n) = W_3^T \sigma(W_4 i_m + W_5 i_n) + M^{fw/bw}_{m,n}
\end{equation}
where the forward and backward matrices are defined as:
\begin{equation}
M_{m,n}^{fw} = {
    \left\{
        \begin{array}{rcl}
        -|d_{m,n}|,  &  m<n \\
        -\infty, & otherwise
        \end{array}
    \right\}
}
\end{equation}

\begin{equation}
M_{m,n}^{bw} = {
    \left\{
        \begin{array}{rcl}
        -|d_{m,n}|,  &  m>n \\
        -\infty, & otherwise
        \end{array}
    \right\}
}
\end{equation}
where $d_{m,n} = exp(|m,n|)$, $|m,n| = 1$ if $i_m,i_n$ are adjacent, and so on. Taking into account the relative position of the two elements, the positional deviation between $i_m$ and $i_n$, $|d_{m,n}|$, are exploited. Note that there are two independent processes containing forward procedure and backward procedure. In the forward encoding process, $M_{m,n}^{fw}$ equals to $-|d_{m,n}|$ when the element $i_m$ is earlier than element $i_n$, $\infty$ otherwise. And vice versa in the backward encoding process.
%For simplicity, we only describe the one-way process, and the reverse is the same.

The attention score between $i_m$ and $i_n$ is defined as:
\begin{equation}
{
    \begin{aligned}
    att.^{fw/bw}(i_m,i_n)= \frac{e^{[cor.^{fw/bw}(i_m,i_n)]}} {\sum_{n=1}^{|T|} e^{[cor.^{fw/bw}(i_m,i_n)]}}
    \end{aligned}
}
\end{equation}

After obtaining attention scores over all elements, the output for $i_n$ is defined as:
\begin{equation}
i^{fw}_n = \sum_{m=1}^{T} att.^{fw}(i_m,i_n) \odot i_m
\end{equation}

\begin{equation}
i^{bw}_n = \sum_{m=1}^{T} att.^{bw}(i_m,i_n) \odot i_m
\end{equation}

% \begin{equation}
% i^{bw}_n = \sum_{n=1}^{L} a^{bw}_{mn} \odot i_m
% \end{equation}

Then we use a fully-connected layer to map the combination of 
$E_u^{fw}=\{\textbf{i}_1^{fw},\textbf{i}_2^{fw},...,\textbf{i}_{T}^{fw} \} \in \mathcal{R}^{{T \times d}}$ and
$E_u^{bw}=\{\textbf{i}_1^{bw},\textbf{i}_2^{bw},...,\textbf{i}_{T}^{bw} \} \in \mathcal{R}^{{T \times d}}$ 
to vector 
$ \textbf{s}_u \in \mathcal{R}^d $: 
\begin{equation}
    \textbf{s}_u = ReLU(W_6[E_u^{fw},E_u^{bw}] + b_3)
\end{equation}
where [,] is a concatenation. $\textbf{s}_u$ is the current historical sequence representation, i.e., the \textit{user preference embedding}, which combines each independent behaviors and considers the dynamic dependency between them, can be used to predict probability of user $u$ engaging item $i$ through a predicting function $F$ , which can be inner product or H-layer MLP:
\begin{equation}
p_{u,i} = \sigma(F(\textbf{s}_u, \textbf{i}))
\end{equation}

All the weight matrices $W_x$ and bias $b_x$ mentioned in methodology are learnable parameters. $\sigma (\cdot)$ is the sigmoid function.

% The self-attention module consists of two parts, one is self-attention matrix for attention score learning, the other is position encoding matrices for incorporating temporal information into sequence encoding, which is inspired by ~\cite{huang2018csan}.
% Two forward and backward Position Encoding Matrices are combined with self-attention architecture, named ``\textit{masked self-attention}'', to preserve the temporal information for sequence modeling.
% The input of self-attention layer is the sequence $\mathcal{B}_u=\{i_{t},t = 1,2,...,T\}$ of user $u$,

\subsection{Meta Learner}

metaCSR aims to excavate the common patterns for recommendation by reformulating cold-start recommendation as a few-shot learning problem and promote generalization ability of the trained model. Gradient (or optimization) based meta-learning has recently emerged as an effective approach for few-shot learning. We extend MAML into the metaCSR framework. The full algorithm is outlined in Algorithm \ref{alg:metaCSR}. 

In the \textbf{\textit{meta-train phase}}, let $p(\mathcal{T})$ denotes the distribution over all tasks. Each task $\mathcal{T}_i$ contains $N$ users, each users provides $K_1$ behavioral sequences as the support set $D_S$, and $K_2$ behavioral sequences as the query set $D_Q$. Each item of the sequences in $D_S$ and $D_Q$ is obtained by representation learning of the Diffusion Representer module. Then the sequences are sent to the Sequential Recommender to be encoded as the users' preference embedding. We consider the model represented by a parametrized function $f_{\theta_1,\theta_2}$ with parameters $\theta_1$ of Diffusion Representer and parameters $\theta_2$ of Sequential Recommender. $\theta_1$ and $\theta_2$ are randomly initialized at the first.

In the sequential recommendation task, the ultimate goal is to rank the ground-truth item $j$ higher than all other items $j^{-}$ ($j^{-} \in \{\mathcal{J}^{-}=\mathcal{I}\ w/o\ j\}$), which is defined as follows:
\begin{equation}
\mathcal{L} = \sum_u \sum_j \sum_{j^{-} \in \mathcal{J}^{-}} -log [p_{u,j} - {p_{u,j^{-}}]}
\end{equation}

In the \textbf{\textit{inner loop optimization}} procedure, we update the task-specific model, i.e., the Sequential Recommender, by one or more gradient descent updates using $D_S$:
\begin{equation}
    \mathop{min}\limits_{\theta_2} \sum_{\mathcal{T}_i\sim p(\mathcal{T})} \mathcal{L}_{\mathcal{T}_i}(f_{\theta_1,\theta_2})
\end{equation}
The inner loop optimization across task is performed via SGD, such that the model parameters $\theta_2$ become $\theta_2^{'}$:
\begin{equation}
    \theta_2^{'} = \theta_2 - \alpha \nabla_{\theta_2} \mathcal{L}_{\mathcal{T}_i}(f_{\theta_1,\theta_2})
\end{equation}
The inner loop works for modeling the personalization of users to obtain personalized user preference embedding, in which we do not update the representation of users and items in Diffusion Representer, because the users'/items' representations are commonly shared.

In the \textbf{\textit{outer loop optimization}} procedure, we update all parameters by optimizing the performance $f_{\theta_1,\theta_2^{'}}$ using $D_Q$:
\begin{equation}
    \mathop{min}\limits_{\theta_!, \theta_2} \sum_{\mathcal{T}_i\sim p(\mathcal{T})} \mathcal{L}_{\mathcal{T}_i}(f_{\theta_1,\theta_2^{'}})
    = \sum_{\mathcal{T}_i\sim p(\mathcal{T})} \mathcal{L}_{\mathcal{T}_i}(f_{\theta_1,\theta_2 - \alpha \nabla_{\theta_2} \mathcal{L}_{\mathcal{T}_i}(f_{\theta_1,\theta_2})})
\end{equation}
The outer loop optimization across task is performed via Adam, such that the model parameters $\theta_1$ and $\theta_2$ are updated as follows:
\begin{equation}
    \theta_1 \gets \theta_1 - \beta \nabla_{\theta_1} \sum_{\mathcal{T}_i\sim p(\mathcal{T})} \mathcal{L}_{\mathcal{T}_i}(f_{\theta_1,\theta_2^{'}})
\end{equation}
\begin{equation}
    \theta_2 \gets \theta_2 - \beta \nabla_{\theta_2} \sum_{\mathcal{T}_i\sim p(\mathcal{T})} \mathcal{L}_{\mathcal{T}_i}(f_{\theta_1,\theta_2^{'}})
\end{equation}
The purpose of this procedure is to learn the desired parameters of the whole model via meta-learning in such a way as to prepare that model for fast adaption to new users after only a few local updates.

In the \textbf{\textit{meta-test phase}}, a few samples of new users are applied to fine-tune the trained model, and then the new model is applied to perform the final recommendations.

\begin{algorithm}
    \caption{metaCSR}
    \label{alg:metaCSR}
    \KwIn{$p(\mathcal{T})$: distribution over tasks;\\
    $\alpha,\beta$: step size hyperparameters;\\
    $\theta_1$: parameters of Diffusion Representer;\\
    $\theta_2$: parameters of Sequential Recommender;}
    % \KwOut{interaction matrix , movie vectors $V$\;}
    Initialize $\theta_1, \theta_2$ randomly\;
    \While{not converged}
    {
      Sample batch of tasks $\mathcal{T}_i \sim p(\mathcal{T})$
      
      \For{all $\mathcal{T}_i$}
      {
        Sample $K_1$ sequences as support set $D_s$ from $\mathcal{T}_i$
        
        Evaluate $\nabla_{\theta_2} \mathcal{L}_{\mathcal{T}_i}(f_{\theta_1,\theta_2})$ using $D_s$
        
        \textbf{Inner loop optimization:}
        
        Compute adapted parameters with gradient descent: 
        
        $\theta_2^{'} = \theta_2 - \alpha \nabla_{\theta_2} \mathcal{L}_{\mathcal{T}_i}(f_{\theta_1,\theta_2})$
        
      }
      
      \textbf{Outer loop optimization:}
      
      Sample $K_2$ sequences as query set $D_Q$ from $\mathcal{T}_i$
      
      Update $\theta_1, \theta_2$ using $D_Q$:
      
      $\theta_1 \gets \theta_1 - \beta \nabla_{\theta_1} \sum_{\mathcal{T}_i\sim p(\mathcal{T})} \mathcal{L}_{\mathcal{T}_i}(f_{\theta_1,\theta_2^{'}})$
      
      $\theta_2 \gets \theta_2 - \beta \nabla_{\theta_2} \sum_{\mathcal{T}_i\sim p(\mathcal{T})} \mathcal{L}_{\mathcal{T}_i}(f_{\theta_1,\theta_2^{'}})$
    }
    %\end{algorithmic}
\end{algorithm}

\section{Experiments}
We mainly focus on the cold-start sequential recommendation task. In this section, we conduct a comprehensive suite of experiments and evaluate the empirical performance of the proposed method in three public real-world datasets, and try to address the following five major research questions (RQs):
\begin{itemize}
    \item \textbf{RQ1}: Can the proposed metaCSR model effectively solve the user's cold-start sequential recommendation problem?
    
    \item \textbf{RQ2}: Can the proposed metaCSR model also perform well in warm-start scenario? 
    
    \item \textbf{RQ3}: How sensitive is the proposed metaCSR model to the amount of training data?
    
    \item \textbf{RQ4}: How much does each module in the framework play a role in the user CSR task?
        
    \item \textbf{RQ5}: Does the proposed metaCSR model learn the common pattern of users' sequential behaviors?
\end{itemize}

\subsection{Settings}
\subsubsection{Datasets.}
We adopt three widely-used datasets to conduct the experiments, including:
% \textbf{MovieLens-1M}\footnote{https://grouplens.org/datasets/movielens/1m/} (We consider the recommendation task targeting for implicit feedback like previous efforts~\cite{yu2014personalized, he2017neural}.). So the threshold of positive rating is set to 4.), 
% \textbf{Last.fm}\footnote{http://millionsongdataset.com/lastfm/} (We intercepte the latest week data for the experiments. Only interactions are used in this study.)
% and 
% \textbf{Amazon-Video}\footnote{http://jmcauley.ucsd.edu/data/amazon/} (We take the "Video" sub-dataset for the experiments.). 

\begin{itemize}
    \item \textbf{MovieLens-1M}\footnote{https://grouplens.org/datasets/movielens/1m/}. It offers the user-item interaction data in movie domain.
    % We filter out the unpopular items which are watched less than 20 times, and inactive users who have fewer than 20 interactions.
    We consider the recommendation task targeting for implicit feedback like previous efforts~\cite{yu2014personalized, sun2018recurrent, he2017neural}. So the threshold of positive rating is set to 4. 
    We define the target value of user-item equals to 1 when user-item interaction is observed and the rating is above threshod, and 0 otherwise.
    
    \item \textbf{Last.fm}\footnote{http://millionsongdataset.com/lastfm/}. It is an official song tag and song similarity dataset of the Million Song Dataset. We intercepted the latest one week data for the experiments. Only interactions are used in this study.
    
    \item \textbf{Amazon-Video}\footnote{http://jmcauley.ucsd.edu/data/amazon/}. The Amazon product dataset, comprising large corpora of reviews and timestamps on various products ~\citep{mcauley2015image}, is notable for its high sparsity and variability. We take "Video" sub-dataset for experiments. 
\end{itemize}

The statistics of datasets are presented in Table ~\ref{tab:dataset}.

\begin{table}
\centering
\renewcommand\arraystretch{1}
  \caption{Statistics of the datasets.}
  \label{tab:dataset}
  \begin{tabular}{c|cccc}
    \toprule
    Datasets & \#Users & \#Items & \#Interactions &Density \\
    \midrule
    MovieLens-1M & 6,040 & 3,952 & 575,281 & 2.41\%\\
    Last.fm-1week & 699 & 45,519 & 59,725 & 0.19\%\\
    Amazon-Video & 368,837 & 19,653 & 482,591 & 0.007\%\\
  \bottomrule
\end{tabular}
\end{table}

\subsubsection{Implementation Details.}
\begin{itemize}
    \item \textbf{General settings.} 
    MovieLens and Last.fm are more evenly distributed, with an average value of interactions about 100, while Amazon-Video presents an obvious long-tail distribution pattern. Therefore, for the MovieLens and Last.fm datasets, we regard 80\% of users with a large number of behaviors as regular users, and the remianing 20\% as new users. 
    In order to simulate real scenarios and evaluate the performance of the model, for all new users, we only keep at most 10 earliest behaviors, and the testing phase is carried out on these few behaviors.
    As for the Amazon-Video dataset, we regard users with behaviors in the range of 2-5 as new users, and users with behaviors greater than 5 as regular users. 
    For each positive sample, we follow \cite{he2017neural} and randomly sample 100 items that the user has not interacted with to generate negative samples that pair it. 
    Given the user's interaction sequence, we random select a short sequence containing $T$ interactions to predict the $(T+1)\textrm{-}th$ interaction in the meta-train phase, where $T\in[2,10]$.
    
    \item \textbf{Meta-learning settings. }
    In the meta-train phase, we use 15way-5shot setting. Every 15 regular users form a meta-task, each user's support set contains 5 samples, and the query set contains 15 samples. 
    In the meta-test phase, we hold out the last behavior of new users to build a cold-start scenario test-set, and the behavioral sequences before the last behavior are used to fine-tune the model to quickly adapt to unseen new users' recommendation tasks. In the warm-start scenario, we hold out the last behavior of regular users to build the final warm-start scenario test-set.
    
    \item \textbf{Implementation details.}
    The whole model is trained in an end-to-end way. We apply a grid search for the learning rate and find $\alpha = 1e^{-4}$ with SGD optimizer and $\beta = 1e^{-2}$ with Adam optimizer is the best. Weight decay = $5e^{-4}$. The batch size is set to 16. All the embedding dimension are set to 128. For the comparative methods, the parameters are set as suggested by the original papers.
    
\end{itemize}

\subsubsection{Evaluate Metrics.}
A variety of widely used evaluation metrics are adopted to evaluate our approach: 
% \textbf{AUC}, Mean Average Precision (\textbf{MAP}), Hit Ratio (\textbf{Hit@N}), and Normalized Discounted Cumulative Gain (\textbf{NDCG@N}).

\begin{itemize}
\item \textbf{AUC}.
We evaluate the ranking performance as practical recommender systems usually generate a ranked list of items for a given user.
AUC~\cite{rendle2009bpr}, the area under the ROC curve, is commonly used for evaluating the quality of a ranking list:
%We report the performance of all methods in terms of the following ranking metrics:
\begin{equation}
\textbf{AUC} = \frac{1}{\mathcal{U}}\sum_{u\in \mathcal{U}}\frac{1}{|J||J^{-}|}\sum_{j\in |J|}\sum_{j\in |J^{-}|}\delta(p_{u,j} > p_{u,j^{-}})
\end{equation}
where $J$ denotes the positive samples set, and $J^{-}$ means negative. $\delta(p_{u,j} > p_{u,j^{-}})$ is an indicator function which returns 1 if $(p_{u,j} > p_{u,j^{-}})$ is true, and 0 otherwise. $p_{u,j}$ is the predicted probability that a user $u\in \mathcal{U}$ may act on $i$ in the test set.
A higher value of AUC indicates better performance for ranking performance.
The floor of AUC from random guess is 0.5 and the best result is 1.

\item \textbf{MAP}. 
Mean Average Precision computes the average value with considering the rank in the sequence of returned items, which is a popular performance measure in information retrieval.
Mean average precision is defined as:
\begin{equation}
MAP = \frac{\sum_{q=1}^{Q}AveP(q)}{Q}
\end{equation}
where $Q$ is the number of queries.

%\item \textbf{MRR} Mean Reciprocal Rank is for evaluating the task that produces a list of possible candidate to a sample of queries, ordered by probability of correctness which refers to rank position of the first relevant item for the query.

\item \textbf{Hit@N}. 
Hit Ratio considers whether the relevant items are retrieved within the top N positions of recommended list. 
It is 1 if any ground-truth items are recommended within the top N items, otherwise 0. 
We compute the mean of all users as the final hit ratio score.

\item \textbf{NDCG@N}. 
Normalized Discounted Cumulative Gain measures the relative orders among positive and negative items within the top N of ranking list. 
It is a standard measure of ranking quality.
\end{itemize}

\subsubsection{Comparison Methods.}
% (1) \textbf{BPR} \citep{rendle2009bpr} is a pairwise ranking framework that takes Matrix Factorization as the underlying predictor. 
% (2) \textbf{LSTM} \citep{zhang2014sequential} units are building unit for the layer of a recurrent neural network (RNN), which are used to capture sequential dependencies and make predictions. 
% (3) \textbf{CSAN} \citep{huang2018csan} is a sequential recommendation algorithm based only on self-attention mechanism, which can speed up the training process and capture user dynamic interests. 
% (4) \textbf{MeLU} \citep{lee2019melu} is a meta-learned user preference estimator for cold-start recommendation, which is a MAML-based algorithm that can identify persanalized preferences.

\begin{itemize}
	\item \textbf{BPR}. 
	Bayesian personalized ranking \citep{rendle2009bpr} is a pairwise ranking framework that takes Matrix Factorization as the underlying predictor. 
	
    \item \textbf{LSTM}
    Long short-term memory (LSTM) units are building units for the layer of a recurrent neural network (RNN), which are used to capture sequential dependencies and make predictions \citep{zhang2014sequential}.
    
	\item \textbf{CSAN}.
	Contextual self-attention network \citep{huang2018csan} is a sequential recommendation algorithm based only on feature-wise self-attention mechanism, which can speed up the training process and capture user dynamic interest.
	
	\item \textbf{MeLU}.
	MeLU \citep{lee2019melu} is a meta-Learned user preference estimator for cold-start recommendation, which is a MAML-based algorithm that can identify personalized preferences.
	
	\item \textbf{MetaCS-L}
	MetaCS-L \citep{bharadhwaj2019meta} is a meta-learning based cold-start recommendation method with a linear regression model as the benchmark model.
	
	\item \textbf{MetaCS-DNN}
	MetaCS-DNN \citep{bharadhwaj2019meta} is a meta-learning based cold-start recommendation method with deep neural network as the benchmark model.

	\item \textbf{SML} 
	SML \citep{zhang2020retrain} is a sequential meta-learning method which offers a general training paradigm, where a neural network-based transfer component can transform the old model to a new model that is tailored for future recommendations.
	
	\item $\rm {\text{\textbf{MetaCF}}_{NGCF}}$
	MetaCF \citep{wei2020fast} aims to learn a suitable model for initializing the adaption with meta-learning, which is applicable to any differentiable CF-based models. We use $\rm \text{MetaCF}_{NGCF}$ as the representative comparison method.

	\item \textbf{metaCSR}.
	Our proposed method.
	
\end{itemize}

\renewcommand{\arraystretch}{1}
\begin{table*}
\small
  \centering
%   \fontsize{6.5}{8}\selectfont
  \caption{The next-one recommendation performance of all the methods across the evaluation metrics AUC and MAP in \textbf{COLD-START scenario}. The best performance is boldfaced; the highest score in baseline is labeled with `$*$'; the percentage in parentheses (+/-\%) represents the relative improvements that metaCSR achieve w.r.t the best baseline.}
  \label{tab:results-cold}
    \begin{tabular}{|p{2.8cm}<{\centering}|p{2.8cm}<{\centering}|p{2.8cm}<{\centering}|p{2.8cm}<{\centering}|}
    \hline
    \multirow{2}{*}{Datasets}&
    \multirow{2}{*}{Methods} &
    \multicolumn{2}{c|}{Evaluation Metrics} \cr\cline{3-4}
    &  &AUC & MAP \cr
    \hline
    \hline
    \multirow{9}{*}{MovieLens-1M} &
    BPR & 0.7355 & 0.2031 \cr\cline{2-4}
    &LSTM & 0.7621 & 0.2119 \cr\cline{2-4}
    &CSAN & 0.7693 & 0.2563* \cr\cline{2-4}
    &MeLU & 0.7391 & 0.1964\cr\cline{2-4}
    &MetaCS-L & 0.7029 & 0.2100\cr\cline{2-4}
    &MetaCS-DNN & 0.7035 & 0.2009\cr\cline{2-4}
    &SML & 0.7621 & 0.1501 \cr\cline{2-4}
    &$\rm \text{MetaCF}_{NGCF}$ & 0.7995* & 0.2506 \cr\cline{2-4}
    &\textbf{metaCSR} & \textbf{0.8623} (+7.85\%) & \textbf{0.2987} (+16.54\%) \cr\cline{1-4}
    
    \multirow{9}{*}{Last.fm-1week} &
    BPR & 0.8333 & 0.2639 \cr\cline{2-4}
    &LSTM & 0.8889 & 0.3903 \cr\cline{2-4}
    &CSAN & 0.9242* & 0.4829 \cr\cline{2-4}
    &MeLU & 0.8586 & 0.2900 \cr\cline{2-4}
    &MetaCS-L & 0.8485 & 0.3122 \cr\cline{2-4}
    &MetaCS-DNN & 0.8788 & 0.5051* \cr\cline{2-4}
    &SML & 0.9053 & 0.4940\cr\cline{2-4}
    &$\rm \text{MetaCF}_{NGCF}$ & 0.9242* & 0.5005\cr\cline{2-4}
    &\textbf{metaCSR} & \textbf{0.9596} (+3.83\%) & \textbf{0.5394} (+6.79\%) \cr\cline{1-4}

    \multirow{9}{*}{Amazon-Video} &
    BPR & 0.7217 & 0.1859 \cr\cline{2-4}
    &LSTM & 0.7831 & 0.2494 \cr\cline{2-4}
    &CSAN & 0.7791 & 0.2509 \cr\cline{2-4}
    &MeLU & 0.7580 & 0.2461 \cr\cline{2-4}
    &MetaCS-L & 0.7207 & 0.2419 \cr\cline{2-4}
    &MetaCS-DNN & 0.7213 & 0.2566*\cr\cline{2-4}
    &SML & 0.7997 & 0.2444 \cr\cline{2-4}
    &$\rm \text{MetaCF}_{NGCF}$ & \textbf{0.8041}* & 0.2530\cr\cline{2-4}
    &\textbf{metaCSR} & 0.8011 (-0.37\%) & \textbf{0.3326} (+29.62\%) \cr\cline{1-4}
    
    \hline
    \end{tabular}
\end{table*}

\begin{figure*}[htbp]
\centering
\subfigure[MovieLens-1M]{
\includegraphics[width=0.9\textwidth]{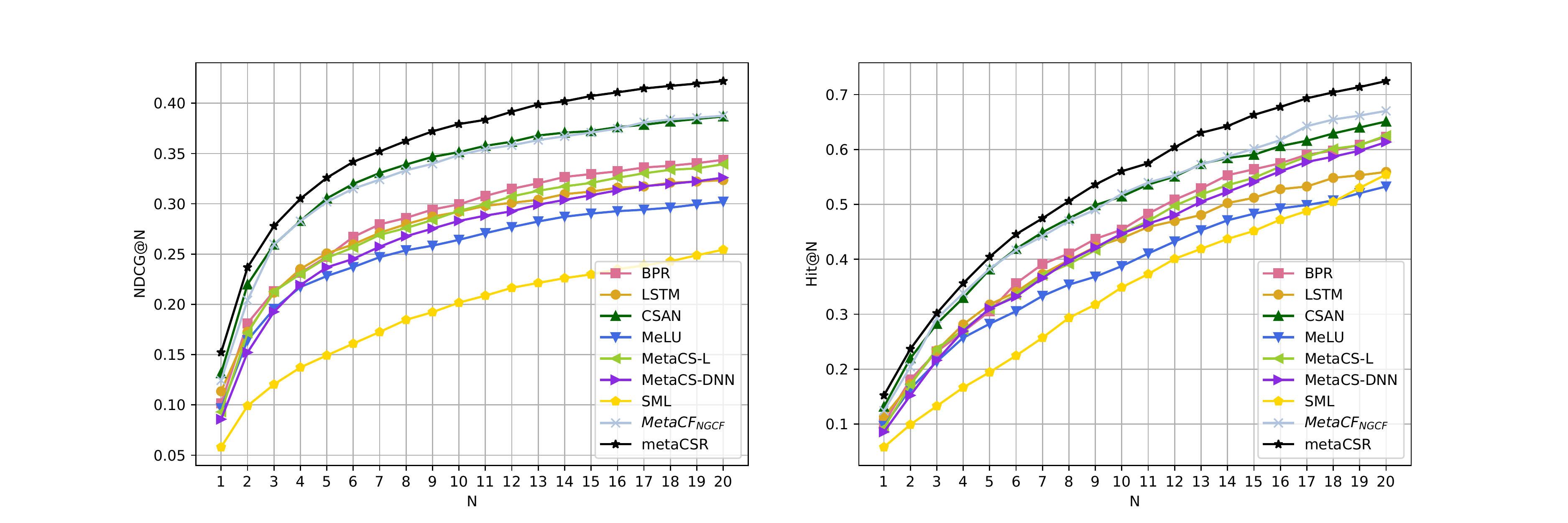}
}
\quad
\subfigure[Last.fm-1week]{
\includegraphics[width=0.9\textwidth]{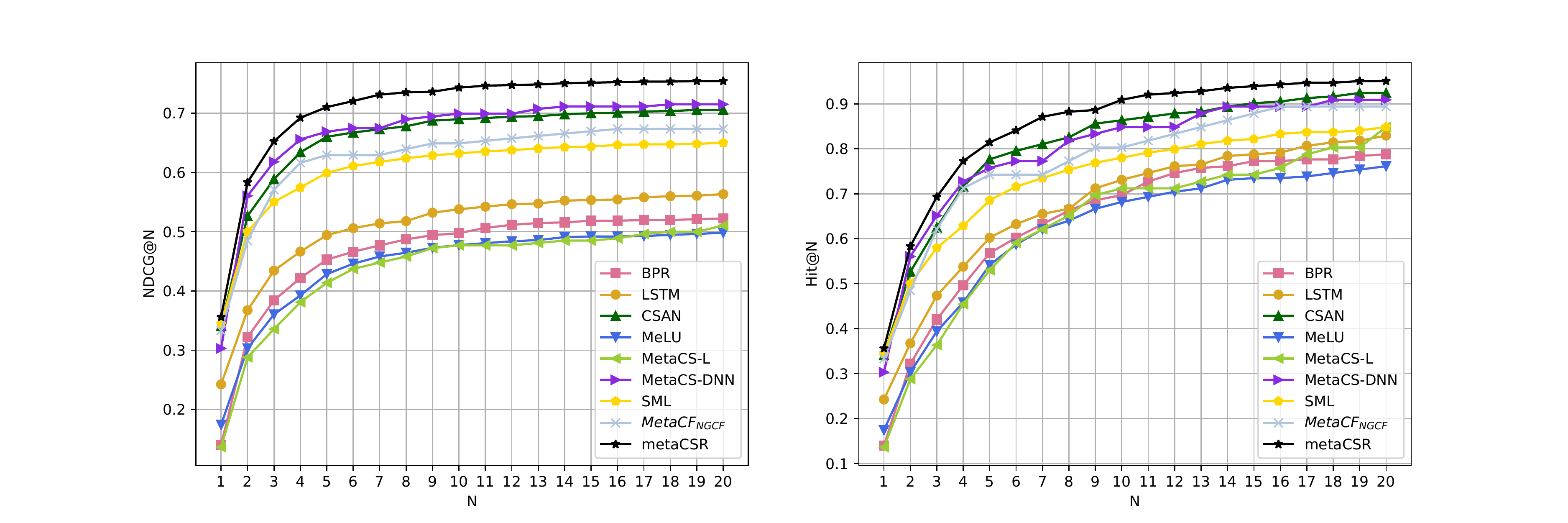}
}
\quad
\subfigure[Amazon-Video]{
\includegraphics[width=0.9\textwidth]{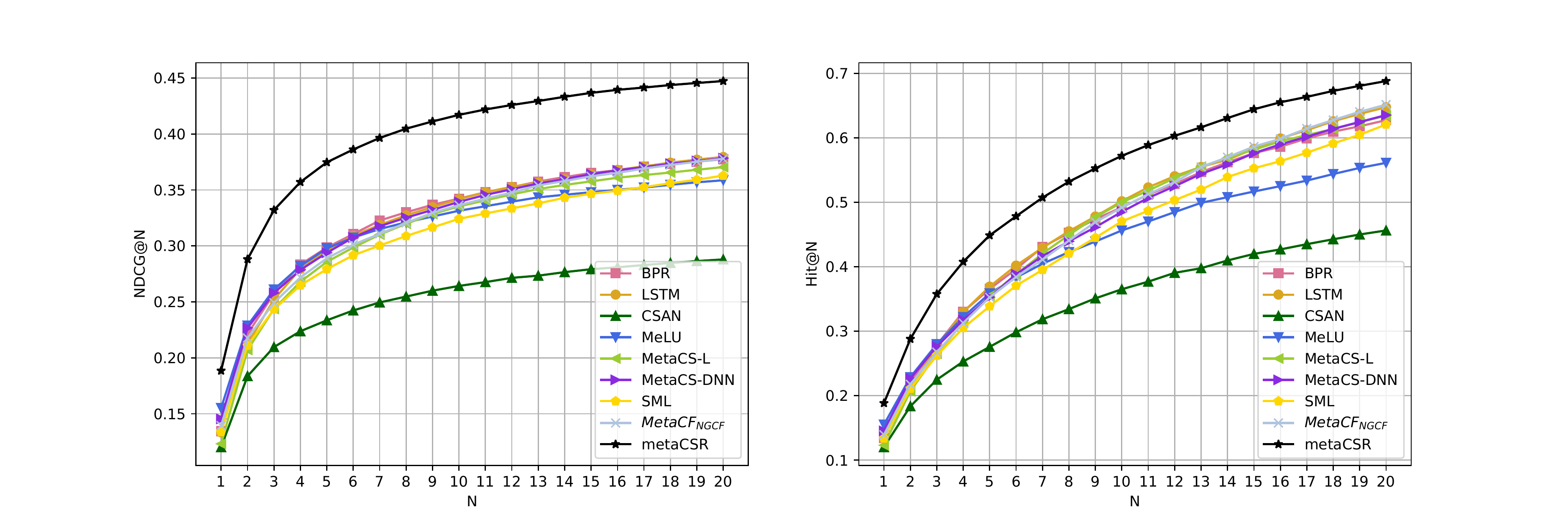}
}
\caption{The next-one recommendation performs of all the methods across the evaluation metrics NDCG@N and Hit@N in \textbf{COLD-START scenario}. $N \in [1, 20]$.}
\label{fig:cold}
\end{figure*}

\subsection{Performance Analysis}

% We mainly focus on the user CSR task. In this section, we evaluate the empirical performance of the proposed method in three public real-world datasets, and try to address the following three major research questions (RQs):

To demonstrate the efficacy of metaCSR, We conduct a comprehensive suite of experiments and evaluate the empirical performance of the proposed method. In this subsection, we perform the analysis of the following five major research questions (RQs):

\subsubsection{RQ1: Can the proposed metaCSR effectively solve the user's cold-start sequential recommendation problem?}
\leavevmode\\

Solving the user's cold-start recommendation problem is the main task of this work, in which new users are not involved in the training of the model as invisible tasks, also called \textit{cold-start scenario} in this paper. In this scenario, new users are held out at the beginning, and the model is fine-tuned only during the testing phase with a very small number of new user's behavioral samples. 
Table \ref{tab:results-cold} shows the next-one recommendation performance of all the methods across the evaluation metrics AUC and MAP on the three datasets in the cold-start scenario, and Fig.~\ref{fig:cold} shows the performance on NDCG@N and Hit@N with N changes. 
It can be seen from the performance under different evaluation metrics:
\begin{enumerate}[(1)]
    \item BPR performs poor than other comparison methods in most cases, because it lacks the ability to model sequential dependencies so that they cannot obtain good performance on sequential recommendation tasks. The following four meta-learning based methods perform better than BPR due to the meta-learning framework they adopt.
    
    \item The sequence modeling methods like CSAN achieves good results in MovieLens-1M and Last.fm-1week datasets, even better than three meta-learning based methods, MeLU, MetaCS-L and MetaCS-DNN, because this method can model sequential patterns, which is beneficial to sequential recommendation tasks. CSAN performs better than LSTM in most cases, due to the fine-grained feature-wise self-attention architecture, which is able to capture the contextual dependencies of sequential behaviors.
    
    \item SML and $\rm \text{MetaCF}_{NGCF}$ perform better than other comparison methods in most cases. $\rm \text{MetaCF}_{NGCF}$ achieves better results because it adopts dynamic sub-graph sampling and incorporates potential interactions for facilitating generalization, which is designed for fast adaptions on new users without relying on user auxiliary information that suitable for our experimental settings.
    
    \item In the cold-start scenario, metaCSR contiguously performs better than comparison methods, and there are significant increases compared to the best baseline results. It demonstrates that our proposed metaCSR model can effectively solve the user's cold-start sequential recommendation problem.
\end{enumerate}

%-------------AUC/MAP in warm-start scenario-----
\renewcommand{\arraystretch}{1}
\begin{table*}
\small
  \centering
%   \fontsize{6.5}{8}\selectfont
  \caption{The next-one recommendation performance of all the methods across the evaluation metrics AUC and MAP in \textbf{WARM-START scenario}. The best performance is boldfaced; the highest score in baseline is labeled with `$*$'; the percentage in parentheses (+/-\%) represents the relative improvements that metaCSR achieve w.r.t the best baseline.}
  \label{tab:results-warm}
    \begin{tabular}{|p{2.8cm}<{\centering}|p{2.8cm}<{\centering}|p{2.8cm}<{\centering}|p{2.8cm}<{\centering}|}
    \hline
    \multirow{2}{*}{Datasets}&
    \multirow{2}{*}{Methods} &
    \multicolumn{2}{c|}{Evaluation Metrics} \cr\cline{3-4}
    &  &AUC & MAP \cr
    \hline
    \hline
    \multirow{9}{*}{MovieLens-1M} &
    BPR & 0.8054  & 0.2079 \cr\cline{2-4}
    &LSTM & 0.8272 & 0.2253 \cr\cline{2-4}
    &CSAN & 0.8414* & 0.2596* \cr\cline{2-4}
    &MeLU & 0.7722 & 0.1885\cr\cline{2-4}
    &MetaCS-L & 0.7446 & 0.2160\cr\cline{2-4}
    &MetaCS-DNN & 0.7371 & 0.2171 \cr\cline{2-4}
    &SML & 0.7935 & 0.1609 \cr\cline{2-4}
    &$\rm \text{MetaCF}_{NGCF}$ & 0.8347 & 0.2134\cr\cline{2-4}
    &\textbf{metaCSR} & \textbf{0.8589}(+2.08\%) & \textbf{0.2742}(+5.62\%) \cr\cline{1-4}
    
    \multirow{9}{*}{Last.fm-1week} &
    BPR & 0.9318 & 0.5075 \cr\cline{2-4}
    &LSTM & 0.9545 & 0.5455 \cr\cline{2-4}
    &CSAN & 0.9798 & 0.5860 \cr\cline{2-4}
    &MeLU & 0.9848* & 0.7156 \cr\cline{2-4}
    &MetaCS-L & 0.9545 & 0.7880 \cr\cline{2-4}
    &MetaCS-DNN & 0.9545 & 0.8404* \cr\cline{2-4}
    &SML & 0.9583 & 0.8009 \cr\cline{2-4}
    &$\rm \text{MetaCF}_{NGCF}$ & 0.9697 & 0.8230 \cr\cline{2-4}
    &\textbf{metaCSR} & \textbf{0.9886}(+0.39\%) & \textbf{0.9097}(+7.37\%) \cr\cline{1-4}

    \multirow{9}{*}{Amazon-Video} &
    BPR & 0.8107 & 0.3532 \cr\cline{2-4}
    &LSTM & \textbf{0.8432}* & 0.3900 \cr\cline{2-4}
    &CSAN & 0.8375 & \textbf{0.4483}* \cr\cline{2-4}
    &MeLU & 0.7522 & 0.2646 \cr\cline{2-4}
    &MetaCS-L & 0.8301 & 0.2748 \cr\cline{2-4}
    &MetaCS-DNN & 0.8285 & 0.2728 \cr\cline{2-4}
    &SML & 0.8363 & 0.2824 \cr\cline{2-4}
    &$\rm \text{MetaCF}_{NGCF}$ & 0.8427 & 0.2831 \cr\cline{2-4}
    &\textbf{metaCSR} & 0.8419(-0.15\%) & 0.3813(-14.95\%) \cr\cline{1-4}
    
    \hline
    \end{tabular}
\end{table*}

\begin{figure*}[htbp]
\centering
\subfigure[MovieLens-1M]{
\includegraphics[width=0.9\textwidth]{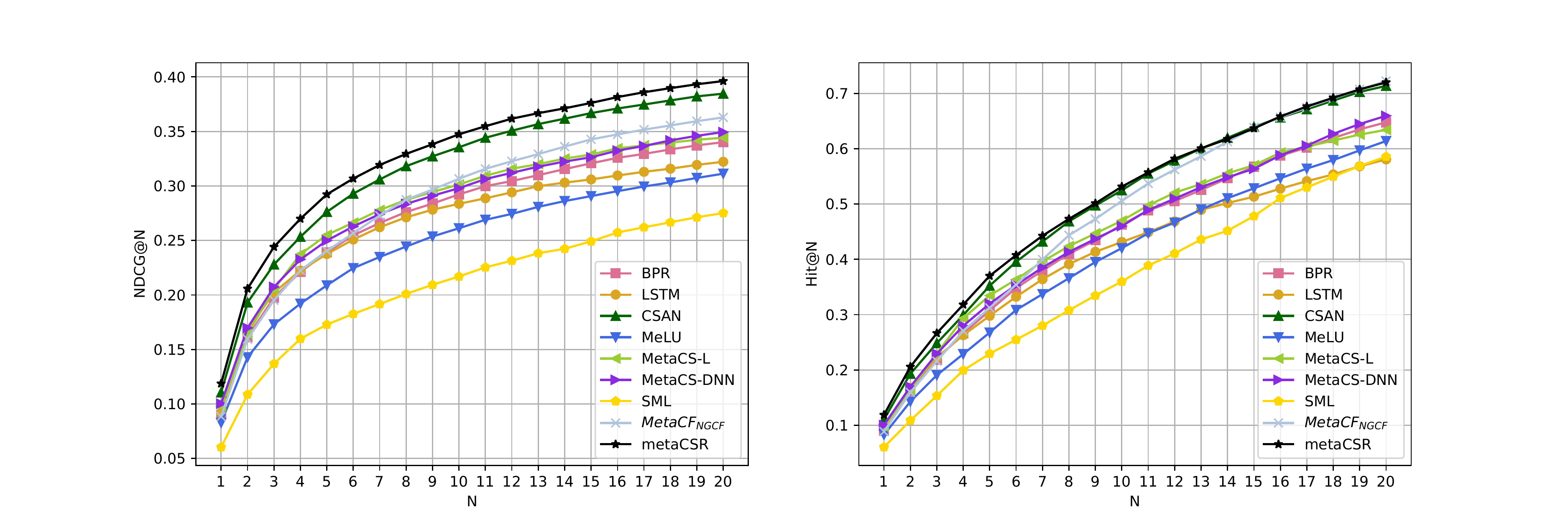}
}
\quad
\subfigure[Last.fm-1week]{
\includegraphics[width=0.9\textwidth]{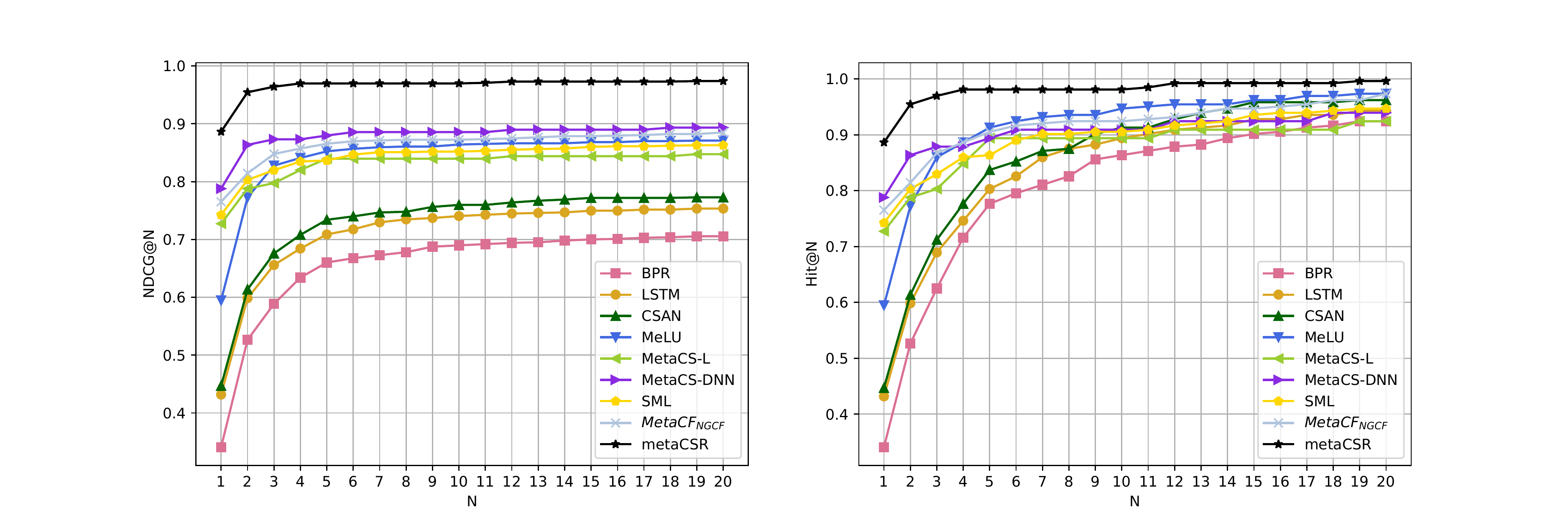}
}
\quad
\subfigure[Amazon-Video]{
\includegraphics[width=0.9\textwidth]{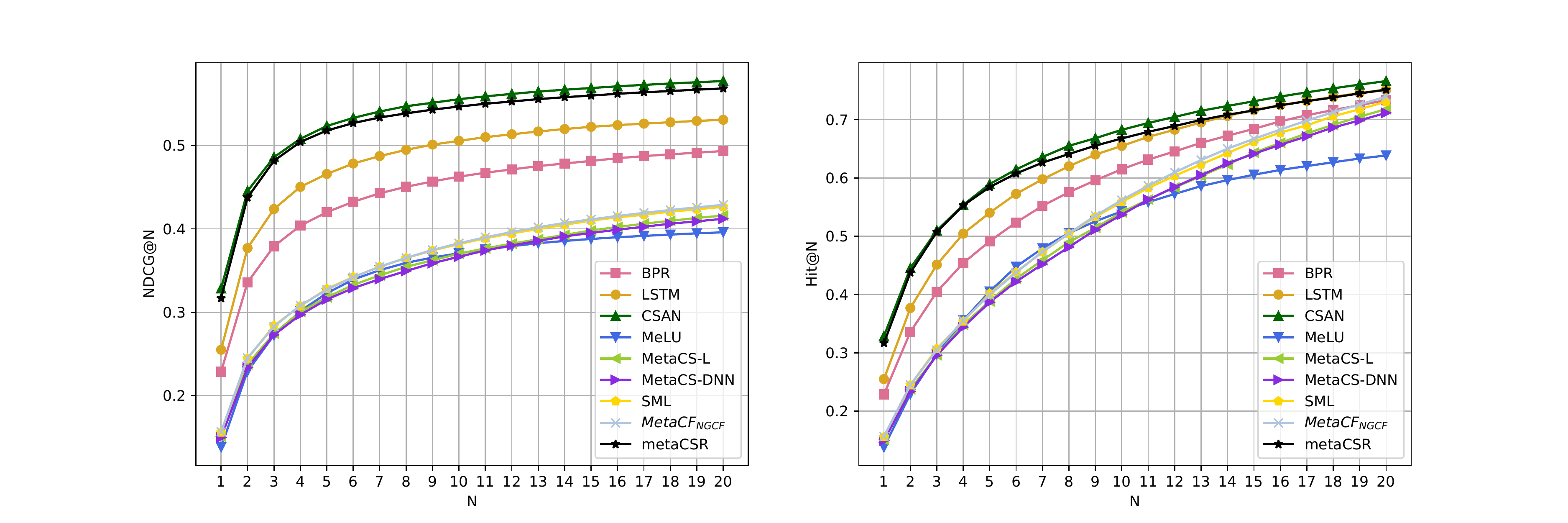}
}
\caption{The next-one recommendation performs of all the methods across the evaluation metrics NDCG@N and Hit@N in \textbf{WARM-START scenario}. $N \in [1, 20]$.}
\label{fig:warm}
\end{figure*}

\begin{figure*}[htbp]
\centering
\subfigure[MovieLens-1M]{
\includegraphics[width=0.4\textwidth]{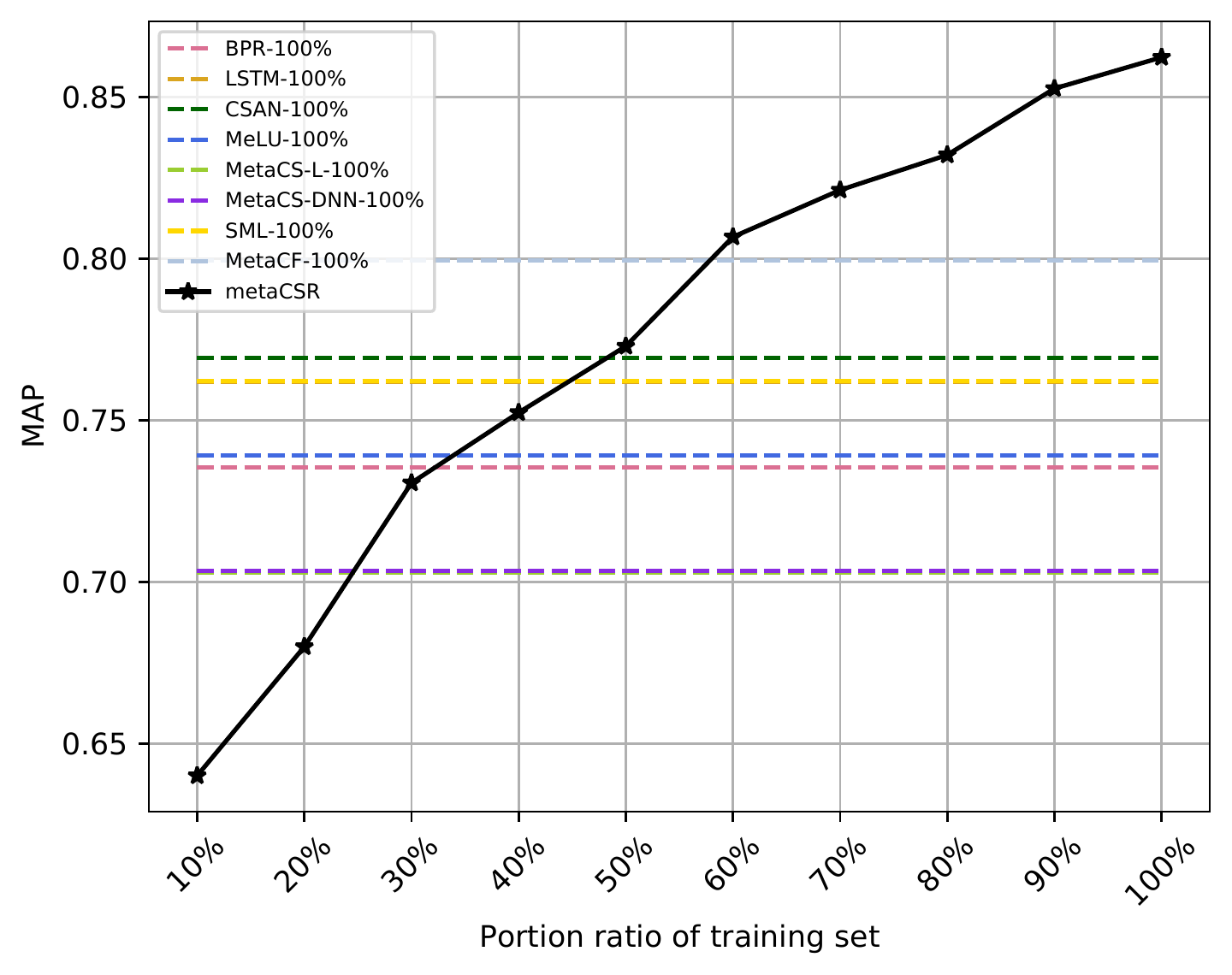}
% \hspace{-5mm}
\includegraphics[width=0.4\textwidth]{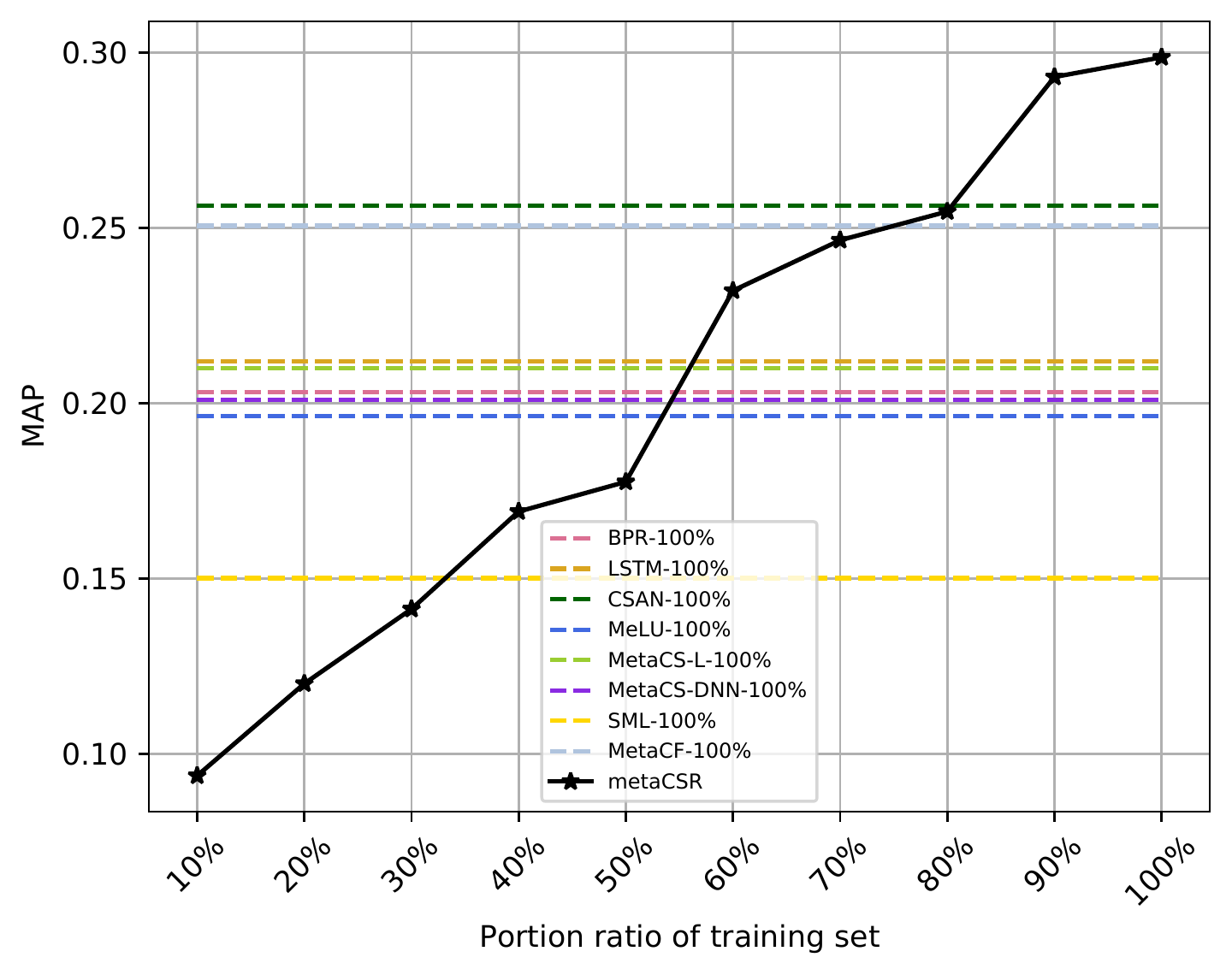}
}
\quad
\subfigure[Last.fm-1week]{
\includegraphics[width=0.4\textwidth]{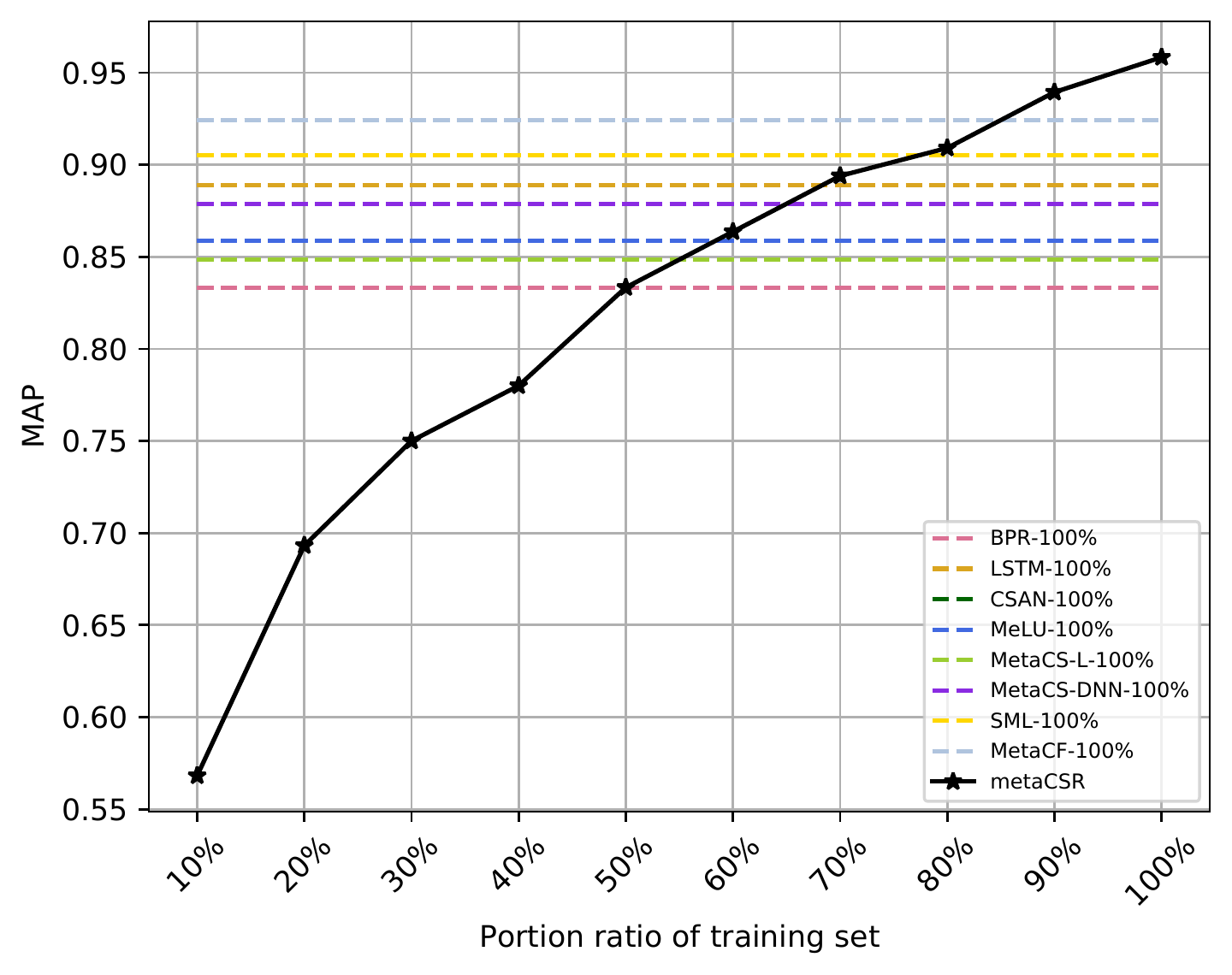}
% \hspace{-5mm}
\includegraphics[width=0.4\textwidth]{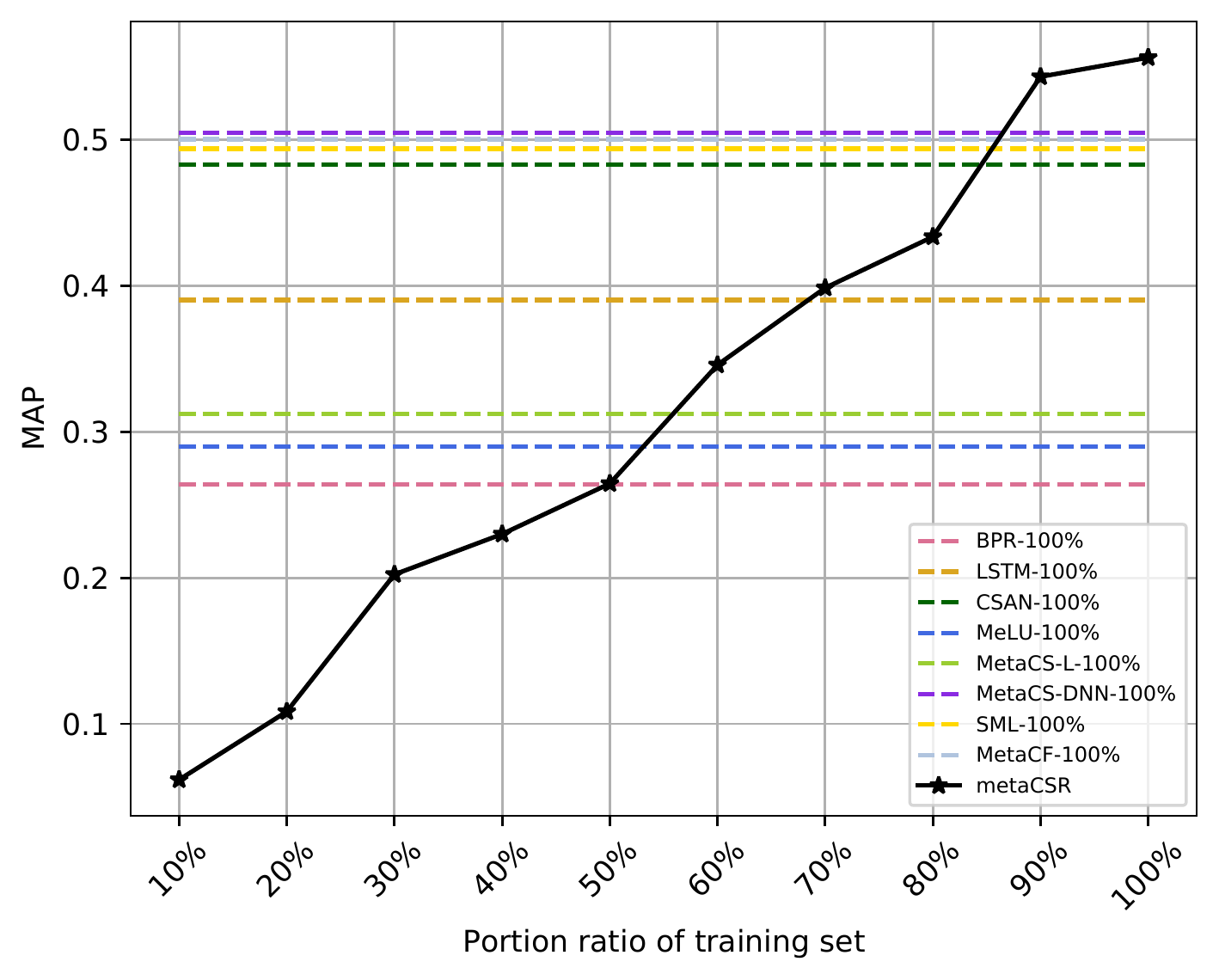}
}
\quad
\subfigure[Amazon-Video]{
\includegraphics[width=0.4\textwidth]{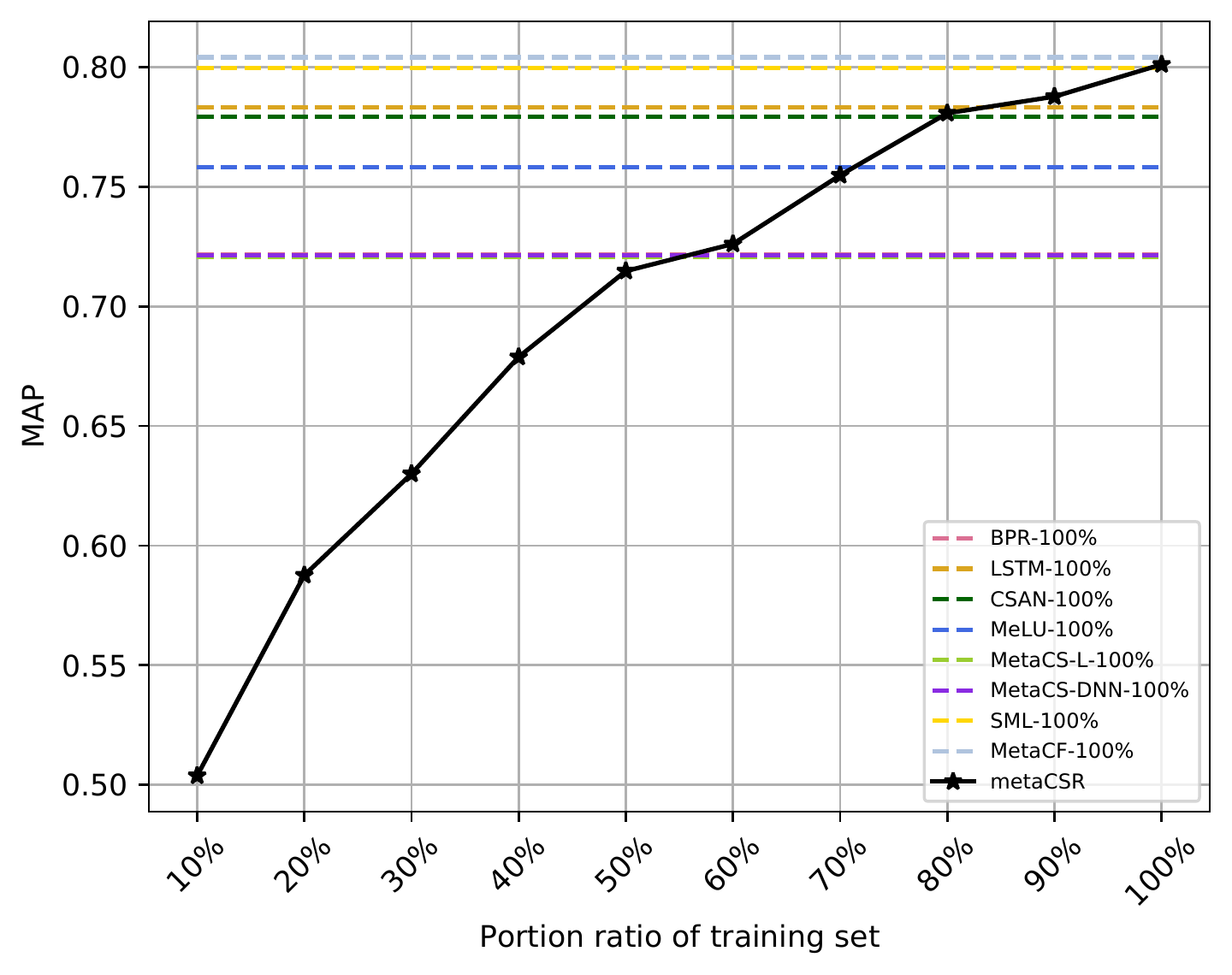}
% \hspace{-5mm}
\includegraphics[width=0.4\textwidth]{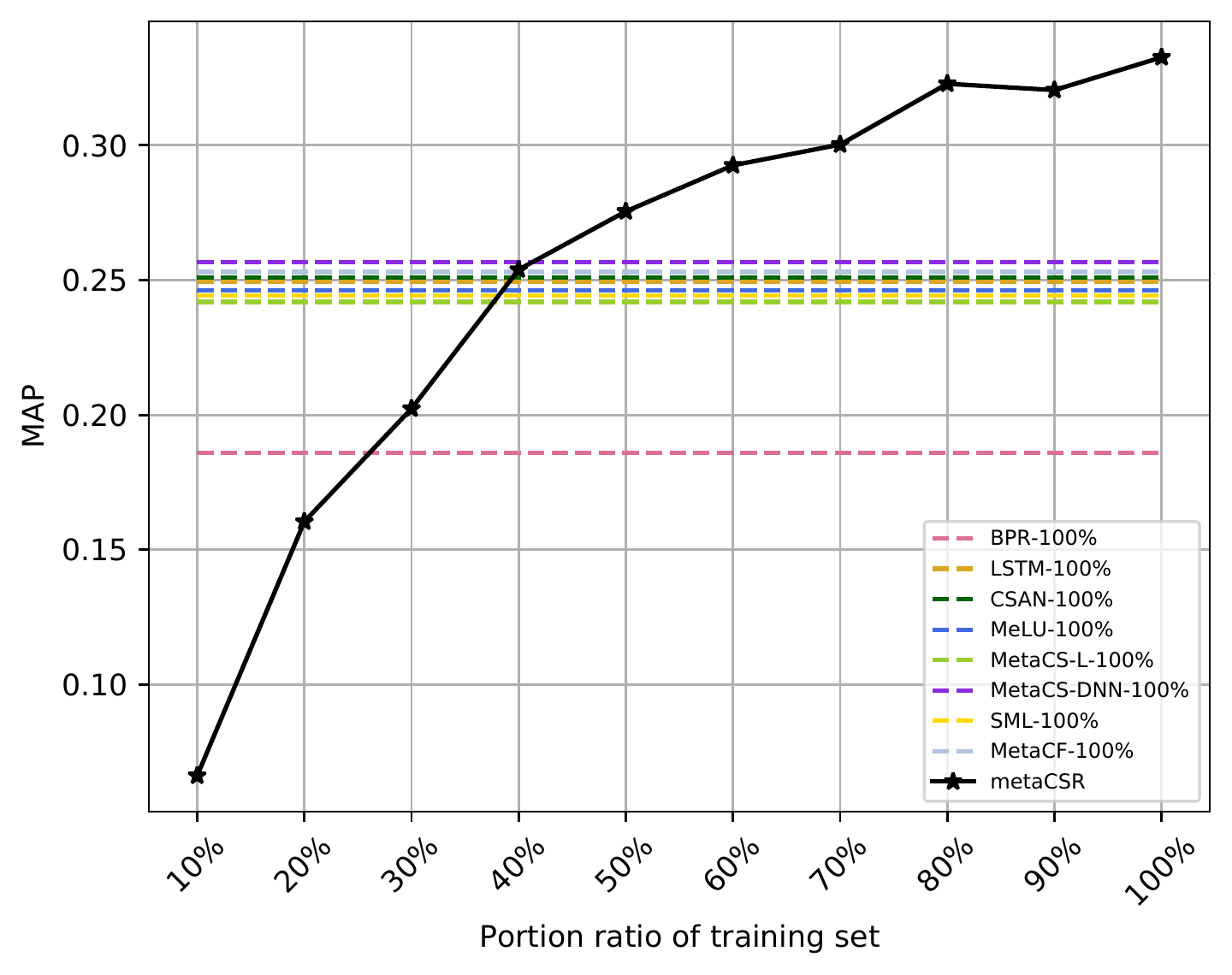}
}
\caption{Sensibility analysis: the trend of AUC and MAP metrics as the amount of training data increases of three datasets.}
\label{fig:sensibility}
\end{figure*}

\subsubsection{RQ2: Can the proposed metaCSR model also perform well in warm-start scenario?}
\leavevmode\\

In addition to the quantitative evaluation of the effect of recommendation to new users in cold-start scenario, we also want to know whether metaCSR is also conducive to improve the performance of regular users. 
Table \ref{tab:results-warm} shows the next-one recommendation performance of all the methods across the evaluation metrics AUC and MAP on the three datasets in warm-start scenario, and Fig.~\ref{fig:warm} shows the performance on NDCG@N and Hit@N with N changes. 
In the \textit{warm-start scenario}, metaCSR achieves the best results in MovieLens-1M and Last.fm-1week datasets with high increasing percentage, which demonstrates the effectiveness of metaCSR performing on regular recommendation task. 

We notice that CSAN gets better results than other methods in Amazon-Video dataset in warm-start scenario. The extreme data sparseness of this dataset makes most models can not well-fit for user interests. 
We assume this is because CSAN is a feature-level attention network that owns the ability of capturing sequential dependencies at a finer-grain level, making up for data sparseness and performing better on regular users. 
In our sequential representer, we only conduct the element-level self-attention network to encode behavior sequence, which is an effective and efficient method in most cases, and it can also be flexibly adjusted under special circumstances. If we face an extremely sparse dataset like Amazon-Video, we can improve performance by using a fine-grained attention network.

Overall, the promising results demonstrate the efficacy of our proposed metaCSR in dealing with user-cold start problem, while maintaining competitive performance in both warm-start and cold-start recommendation scenarios.

\subsubsection{RQ3: How sensitive is the proposed metaCSR model to the amount of training data?}
\leavevmode\\

In order to verify the ability of our model to transfer knowledge and adapt quickly to new tasks, and to test the sensitivity of the model to the amount of training data, we conduct a series of experiments.
We randomly divide the training data into 10 parts according to the number of users, start training the model from 10\% of the data, and conduct a cold start recommendation test on new users. After every 10
\% increase in training data, the model is trained to convergence and the test results are recorded. 

Fig.~\ref{fig:sensibility} shows the trend of AUC and MAP metrics as the amount of training data increases of three datasets. Among them, the dotted line represents the results of the eight baseline methods when the amount of training data is 100\%. Different indicators measure different aspects of model performance. Although the values are biased, overall model performance tends to be the same. For all data sets, metaCSR only needs part of the training data, from 25\% to 85\%, to achieve the best performance of other baseline algorithms. This proves that our proposed model can learn enough knowledge to adapt to new tasks on less training data, so as to obtain better performance in cold-start recommendation tasks.

\begin{table*}
\centering
\renewcommand\arraystretch{1}
  \caption{Ablation studies on MovieLens-1M dataset.}
  \label{tab:abalation}
%   \begin{tabular}{p{0.7cm}<{\centering}|p{1.2cm}<{\centering}p{1.5cm}<{\centering}p{1cm}<{\centering}|p{0.9cm}<{\centering}p{0.9cm}<{\centering}}
    \begin{tabular}{c|ccc|cc}
    \toprule
     Index & Diffusion Representer & Sequential Recommender & Meta Learner & AUC & Improve\% \\
    \midrule
     1 & $\times$ & \Checkmark & \Checkmark & 0.8285 & -3.92\%\\
     2 & \Checkmark & $\times$  & \Checkmark & 0.8297 & -3.78\%\\
     3 & \Checkmark & \Checkmark & $\times$ & 0.8056 & -6.58\%\\
     4 & \Checkmark & \Checkmark & \Checkmark &  \textbf{0.8623} & -\\
    
  \bottomrule
\end{tabular}
\end{table*}

\subsubsection{RQ4: How much does each module in the framework play a role in the user CSR task?}
\leavevmode\\

We conduct some ablation experiments of removing each module from metaCSR framework separately for user CSR task on MovieLens-1M dataset. The results are shown in Table \ref{tab:abalation}. There are some observations:

\begin{enumerate}[(1)]
    \item AUC decreases to varying degrees by removing any one module.
    \item The decrease in performance of removing Diffusion Representer (Index 1) indicates that user/item representations can be improved to some extent through the information diffusion on the graph. 
    \item In the absence of other additional side information, the sequential dependencies of user behaviors are particularly important, which is helpful for capturing the user's dynamic interest drifting, so as to achieve better results in the next-one recommendation (Index 2).
    \item Meta Learner is the most important module in metaCSR. After removing Meta Learner, the performance dropped sharply (Index 3), indicating that Meta Learner actually learns some common knowledge from prior tasks and has a good generalization in new tasks. 
\end{enumerate}

More analysis of Meta Learner can be found in RQ5.

\begin{figure*}[!htbp]
    \centering
    \includegraphics[width=0.8\textwidth]{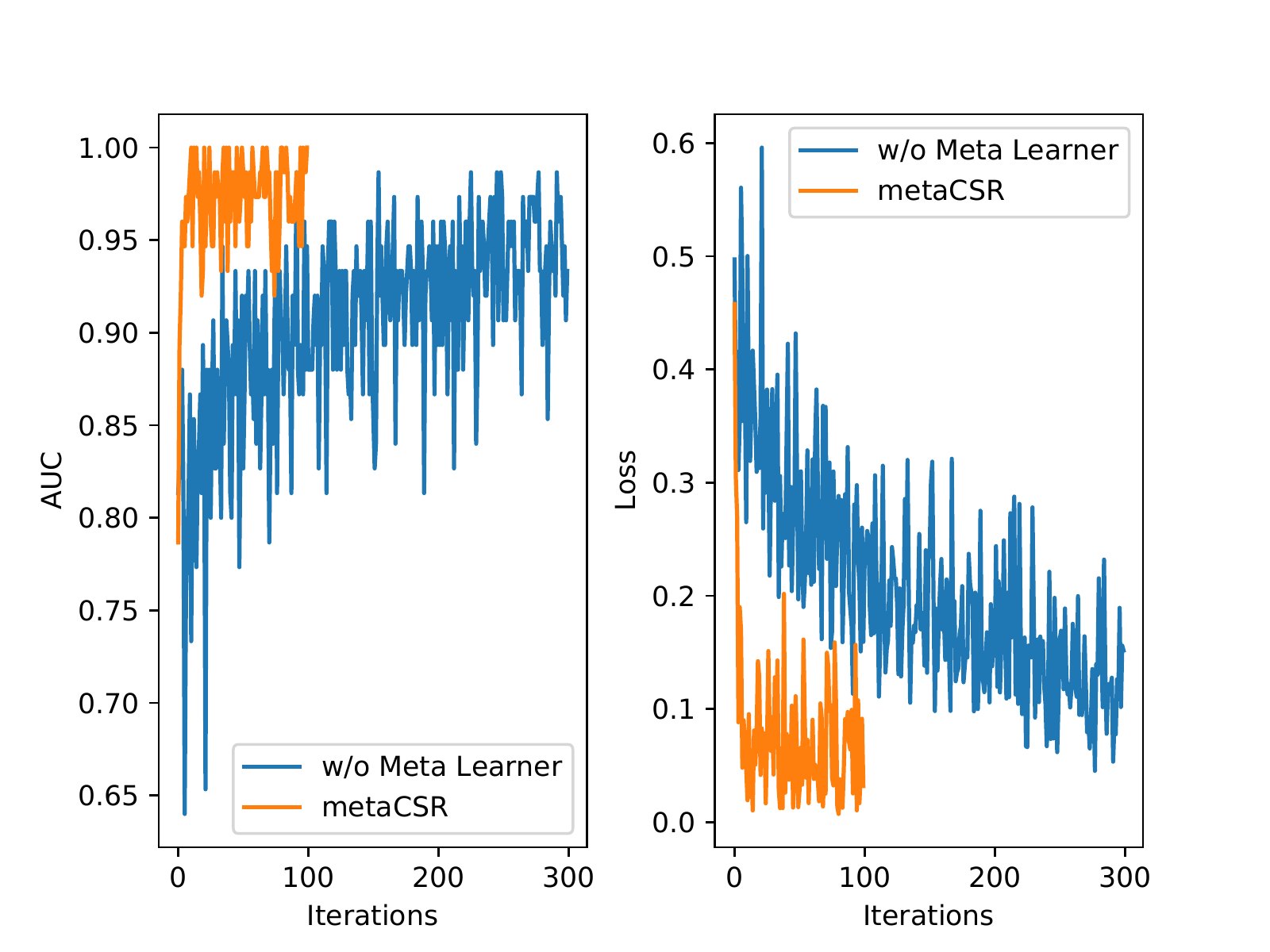}
    \caption{AUC and Loss change with the number of iterations increasing. It is validated on MovieLens-1M dataset.}
    \label{fig:loss}
\end{figure*}  

\subsubsection{RQ5: Does the proposed metaCSR model learn the common patterns of users' behaviors from previous tasks?}
\leavevmode\\

Meta-learning extracts common knowledge from learning different tasks and transfers it for unseen tasks. In addition to the quantitative results as detailed described in RQ1, RQ2, RQ3 and RQ4, we also conduct some visualizations and qualitative analysis of the outputs of the model to verify the efficacy of metaCSR.
The experiments of this section are validated on MovieLens-1M dataset.

\textbf{Faster convergence rate.}

Fig. \ref{fig:loss} shows the changes of AUC and Loss with the number of iterations during meta-test phase in the user CSR task. Compared with ``w/o Meta Learner'' method (removing Meta Learner from metaCSR), metaCSR converge rapidly after a few updates and maintain good performance as the number of iteration increased. It demonstrates that metaCSR extract and propagate transferable knowledge of prior users and learn a good initialization for new users. The excellent efficacy of fast adaptation endows metaCSR algorithms the ability to run online. 

\textbf{Better sample discrimination.}

\begin{figure*}[htbp]
\centering
\subfigure[metaCSR]{
\includegraphics[width=7.2cm]{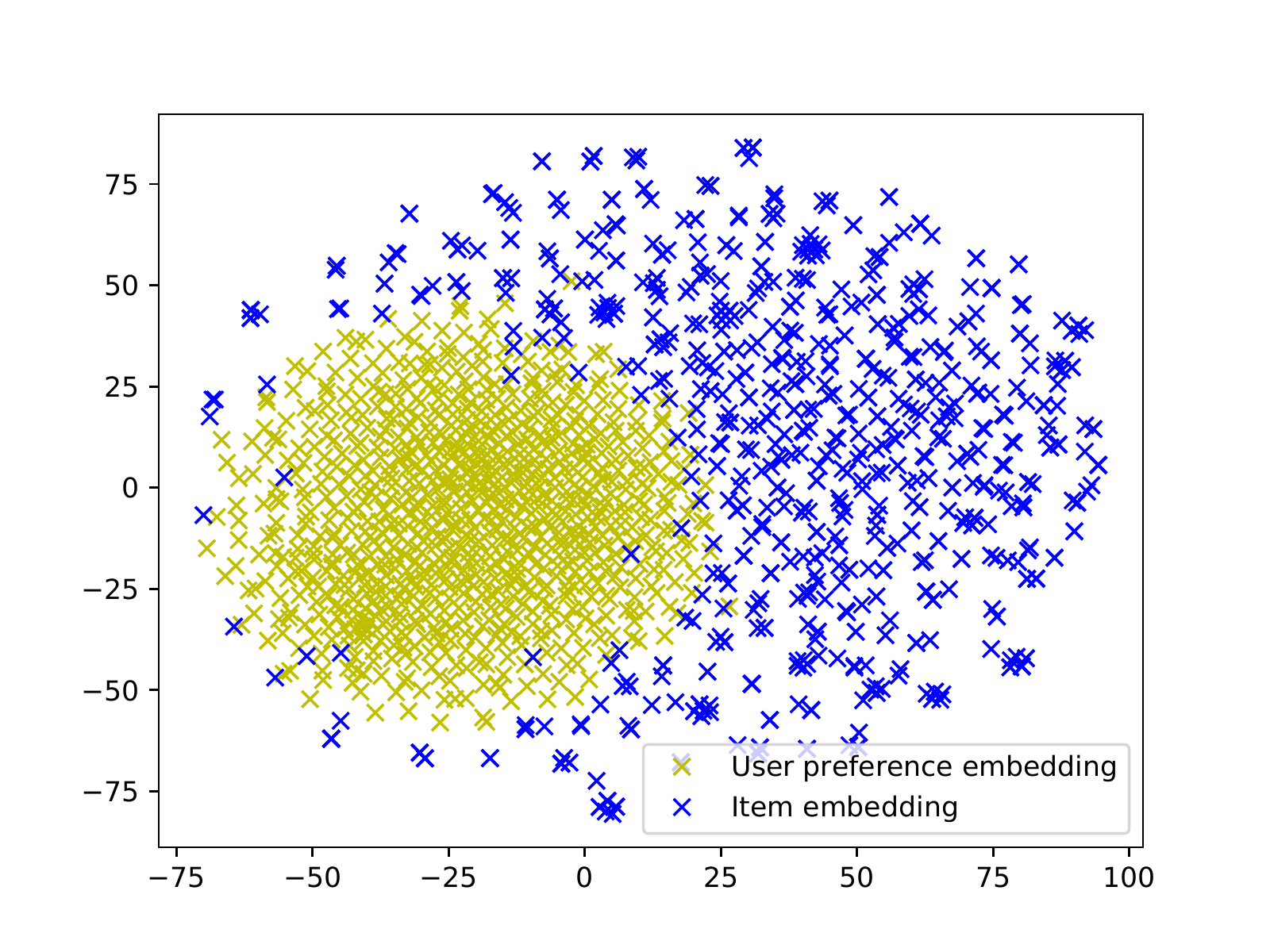}
}\hspace{-13mm}
\quad
\subfigure[w/o Meta Learner]{
\includegraphics[width=7.2cm]{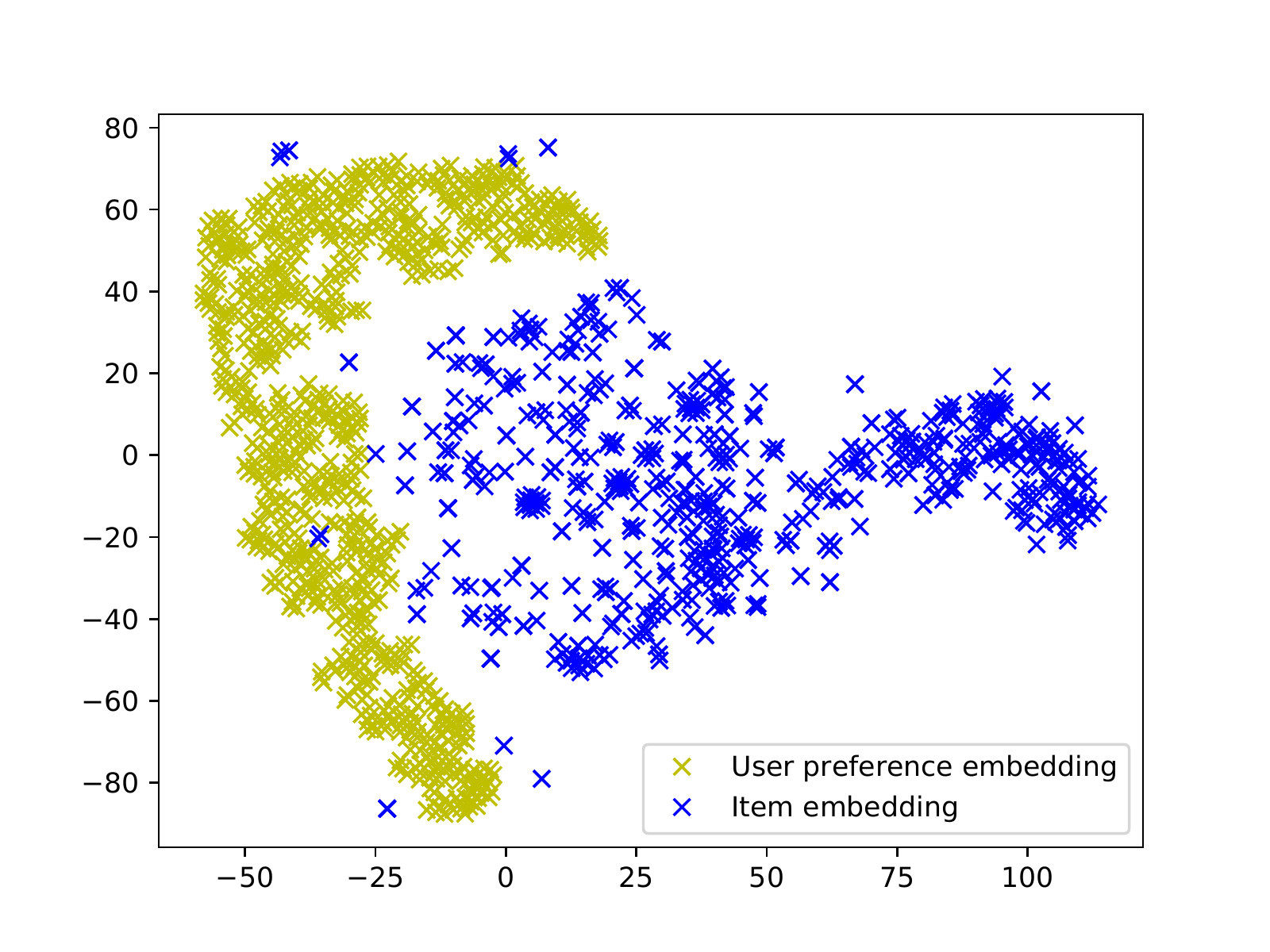}
}
\caption{Visualization of \textbf{NEW} users’ preference embedding and items embedding.}
\label{fig:new}
\end{figure*}

\begin{figure*}[htbp]
\centering
\subfigure[metaCSR]{
\includegraphics[width=7.2cm]{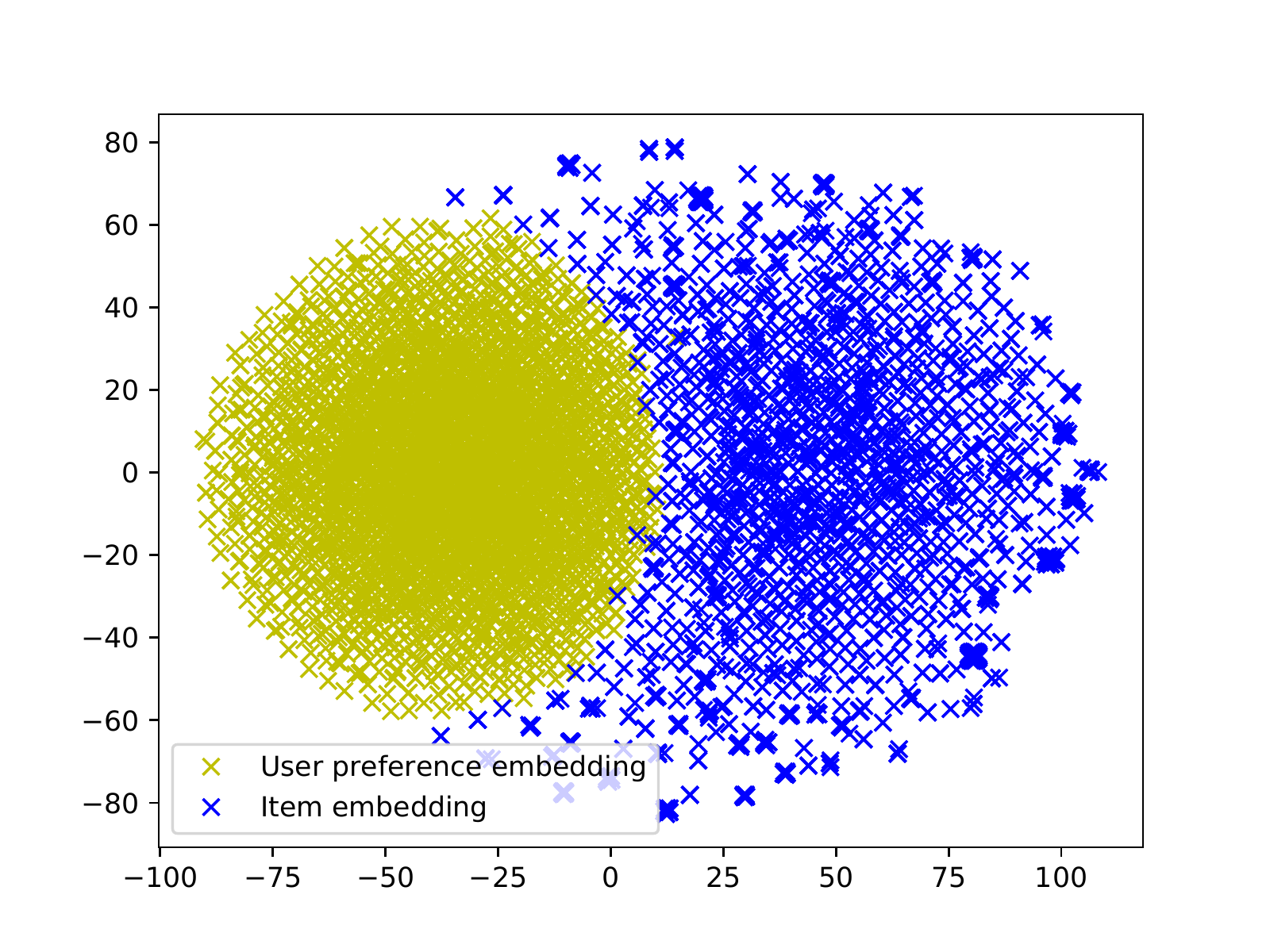}
}\hspace{-13mm}
\quad
\subfigure[w/o Meta Learner]{
\includegraphics[width=7.2cm]{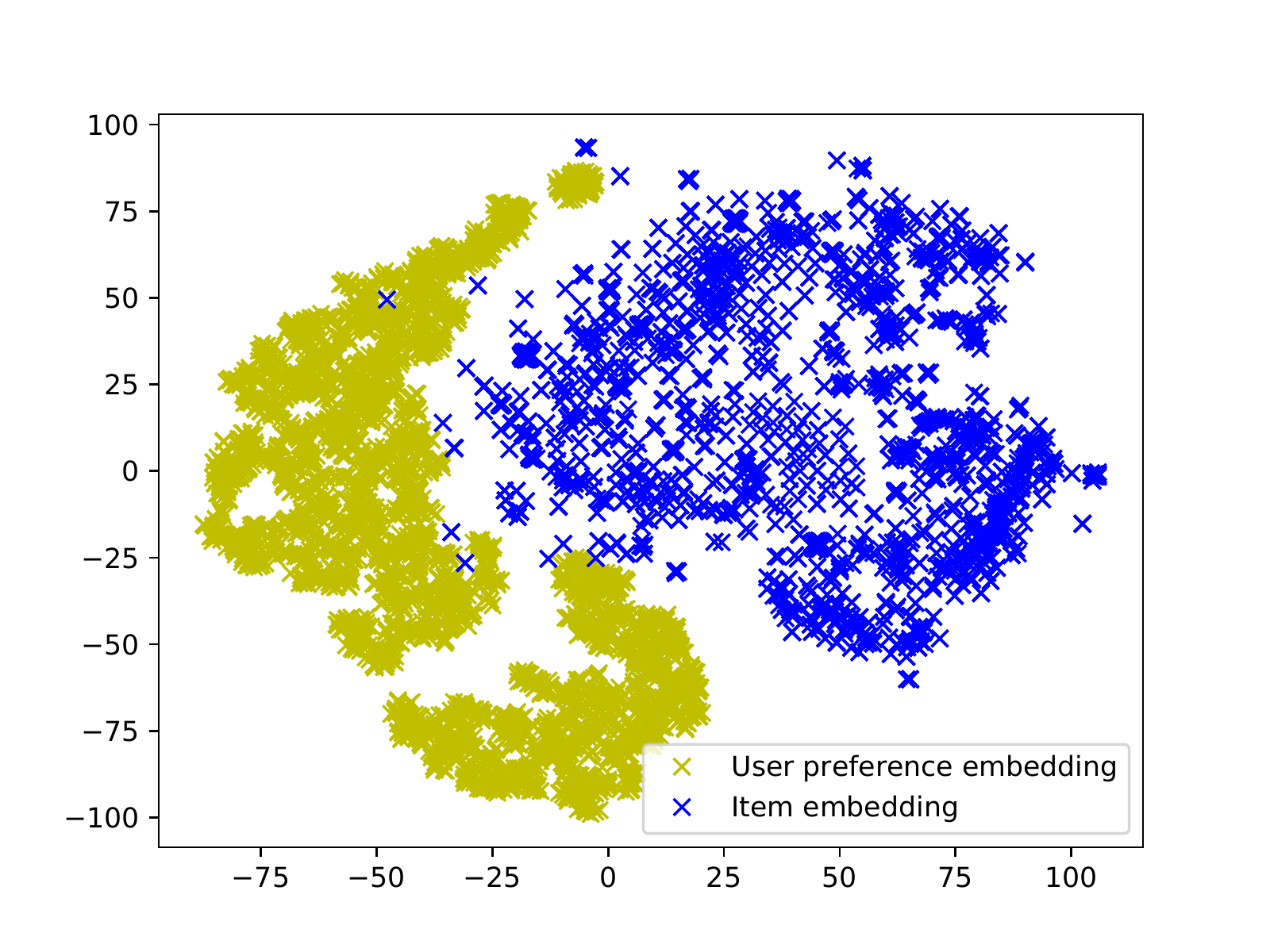}
}
\caption{Visualization of \textbf{REGULAR} users’ preference embedding and items embedding.}
\label{fig:regular}
\end{figure*}

As shown in Fig.\ref{fig:new}-\ref{fig:regular}, we visualize the new/regular users preference embedding (outputs of the model) and items embedding in different situations via t-SNE. Users preference embedding and items embedding in metaCSR (Fig.~\ref{fig:new}-\ref{fig:regular} (a) compared to Fig.~\ref{fig:new}-\ref{fig:regular} (b)), especially on new users (Fig.~\ref{fig:new} (a)), the embeddings have produced a degree of spatial convergence. As model training tends to pull the space of users preference embedding and items embedding closer, so as to achieve better performance when reaching the user-item match during the recommendation process.

\begin{figure*}[!htbp]
    \centering
    \includegraphics[width=1\textwidth]{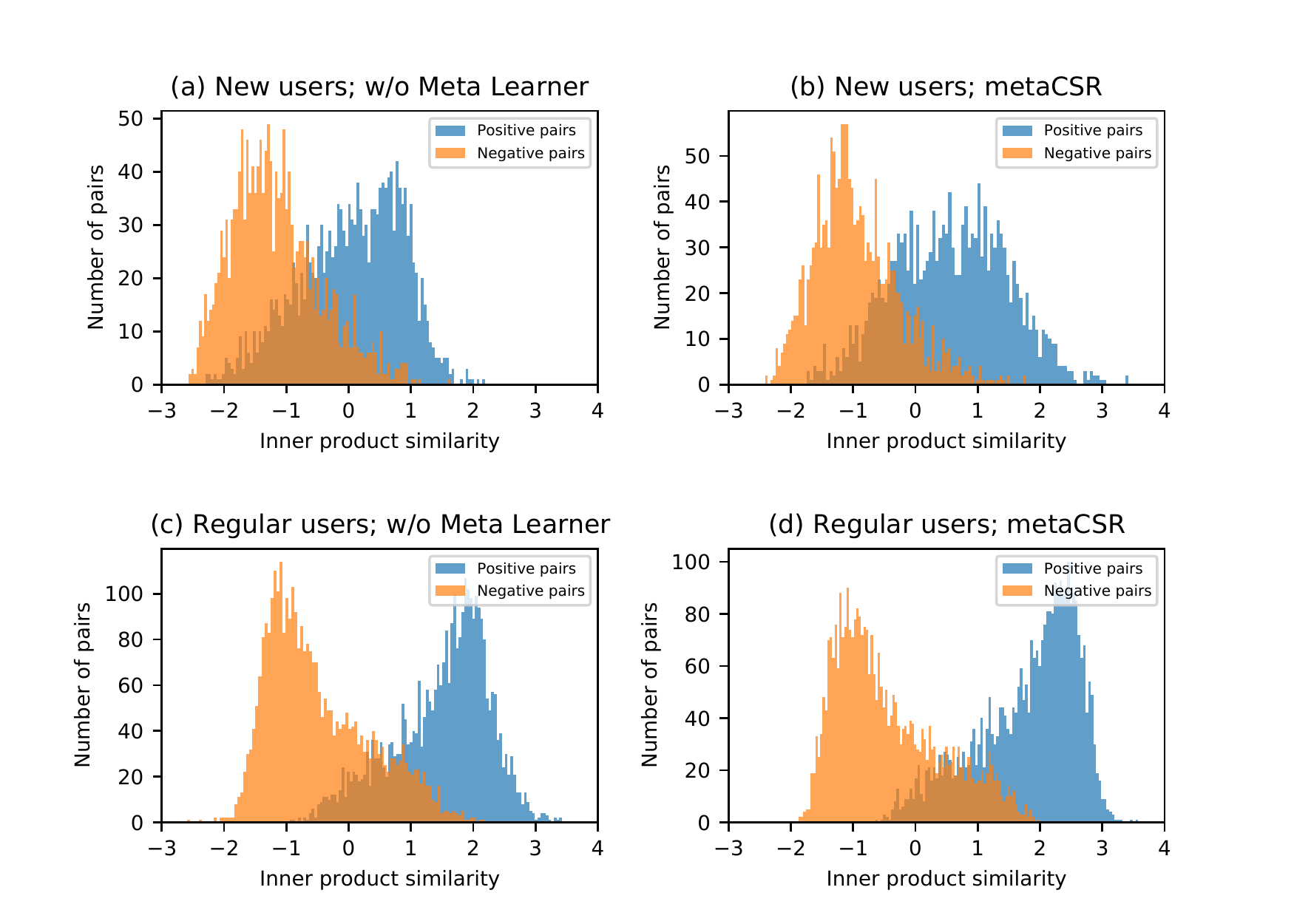}
    \caption{Distributions of inner product similarity of positive pairs and negative pairs in different scenarios. It is validated on MovieLens-1M dataset.}
    \label{fig:distribution}
\end{figure*}

To further intuitively demonstrate the ability of our method to distinguish samples, we visualize the distribution of inner-product similarity of \textit{positive pairs} [user-preference-embedding, positive-item embedding] and \textit{negative pairs} [user-preference-embedding, negative-item embedding] on MovieLens-1M dataset, as shown in Fig. \ref{fig:distribution}.
It is a commonly used and effective way to observe sample discrimination. If the overlap region (histogram intersection) between the two distributions is smaller, the difference between positive and negative samples is greater, and the model can better distinguish between positive and negative samples.
There are some observations:
\begin{enumerate}
    \item The similarity of positive and negative sample pairs corresponding to new users is lower than that of regular users. For example, the similarity of positive pairs of new users under metaCSR model training (Fig. \ref{fig:distribution}(b)) is distributed in the [-2,3] interval, mainly concentrated in the [0-1] interval. The positive sample pair similarity of regular users under the metaCSR model training (Fig. \ref{fig:distribution}(d)) is distributed in the [0,3] interval, mainly concentrated in the [2-3] interval. This is because the model is trained on regular user data, thus the spatial consistency of which is higher.
    
    \item Compared with Fig. \ref{fig:distribution}(a), the overlap region of the positive and negative sample pairs of new users in metaCSR (Fig. \ref{fig:distribution}(b)) is smaller, which means that the difference between the positive and negative samples is greater. Typically, a smaller overlap region indicates better matching performance, since it means more discriminating embeddings are learned. Meanwhile, the overall similarity of the samples in Fig. \ref{fig:distribution}(b) is higher than Fig. \ref{fig:distribution}(a). The above observations show that our method increases the difference between positive and negative sample pairs, and can improve the retrieval accuracy in that sample space. 
\end{enumerate}

\begin{figure*}[htbp]
\centering
\subfigure[metaCSR]{
\includegraphics[width=7.2cm]{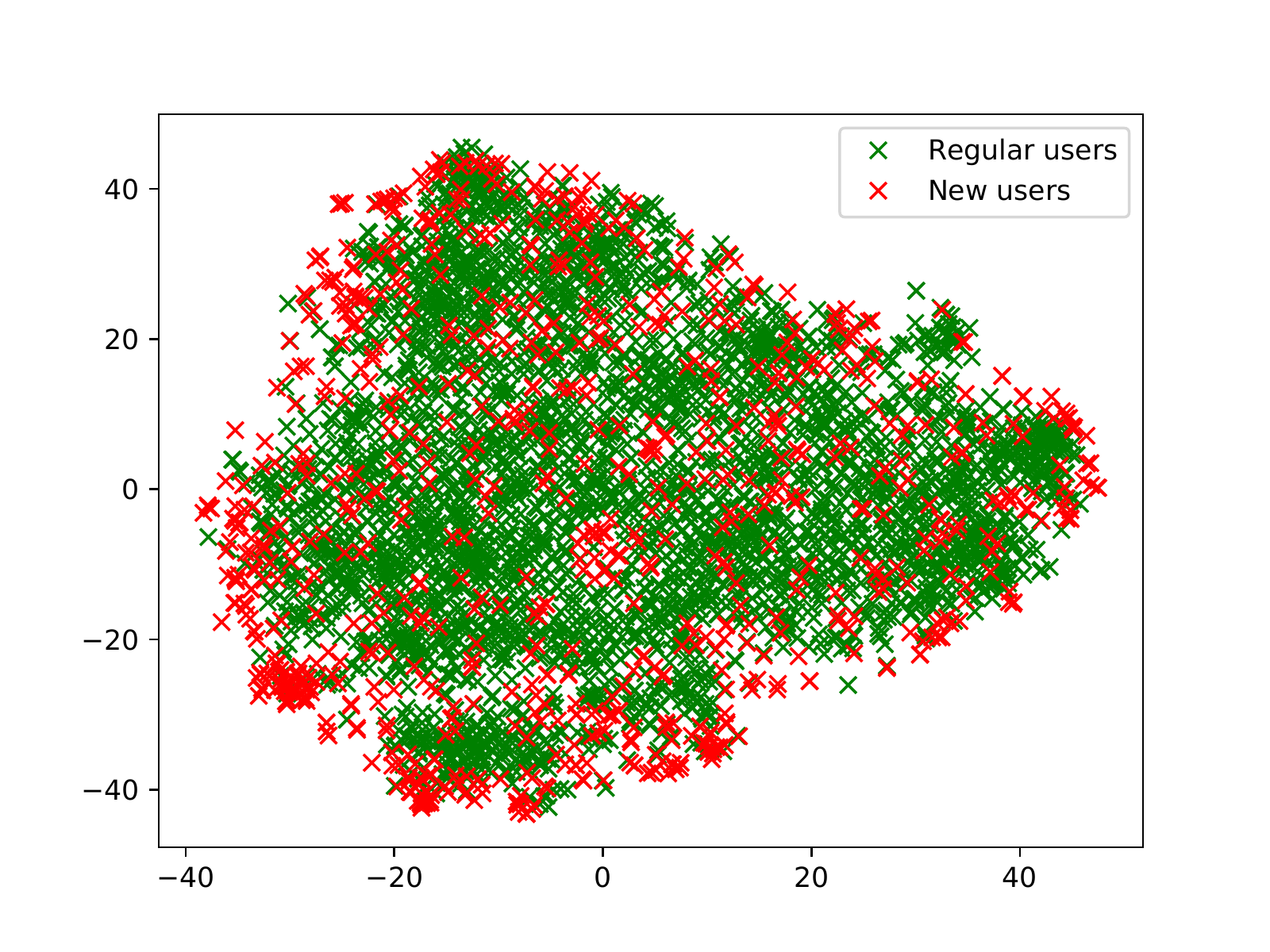}
}\hspace{-13mm}
\quad
\subfigure[w/o Meta Learner]{
\includegraphics[width=7.2cm]{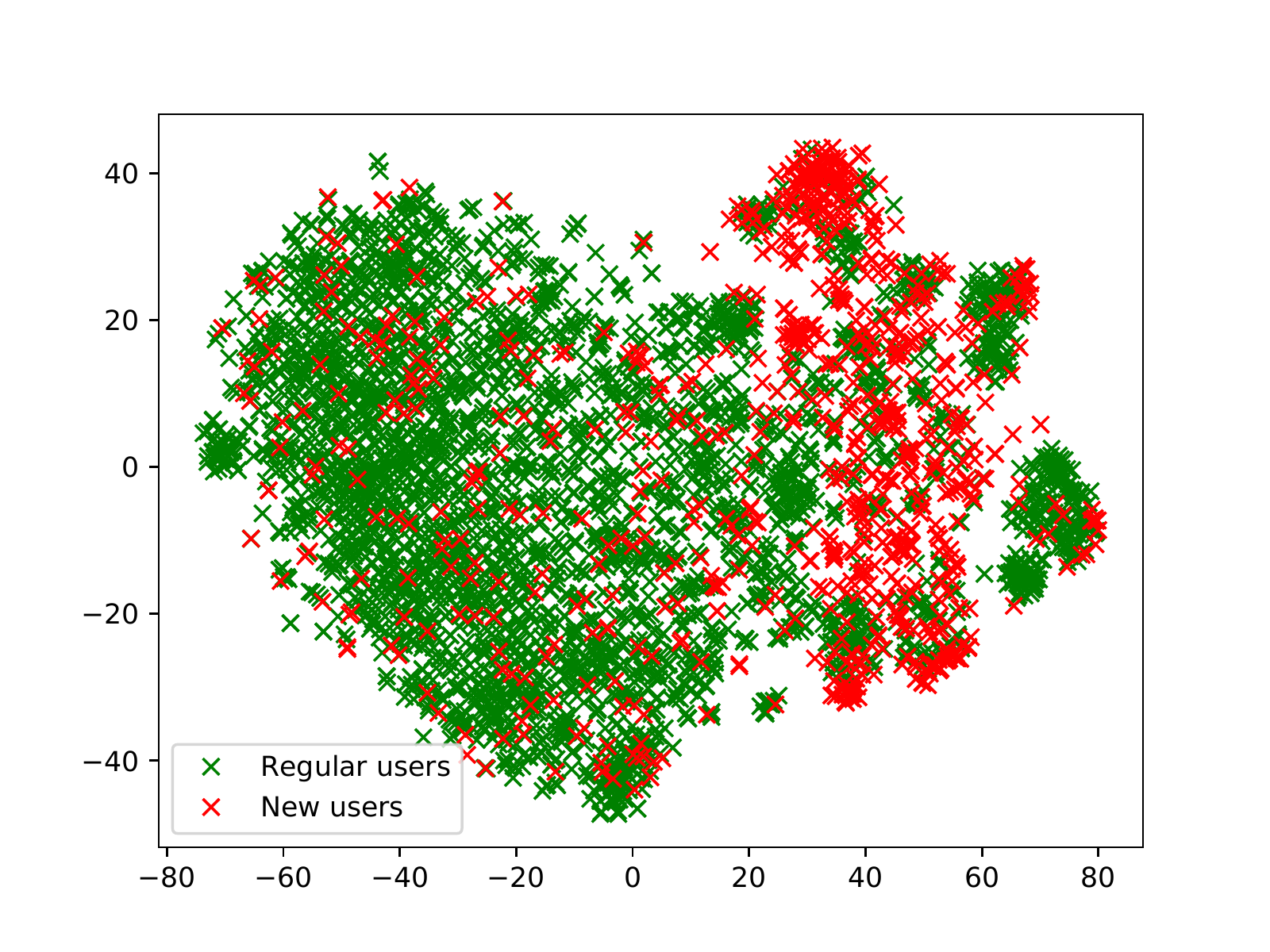}
}
\caption{Visualization of new/regular users’ preference embedding \textbf{WITHOUT fine-tuning}.}
\label{fig:w/o fine-tuning}
\end{figure*}

\begin{figure*}[htbp]
\centering
\subfigure[metaCSR]{
\includegraphics[width=7.2cm]{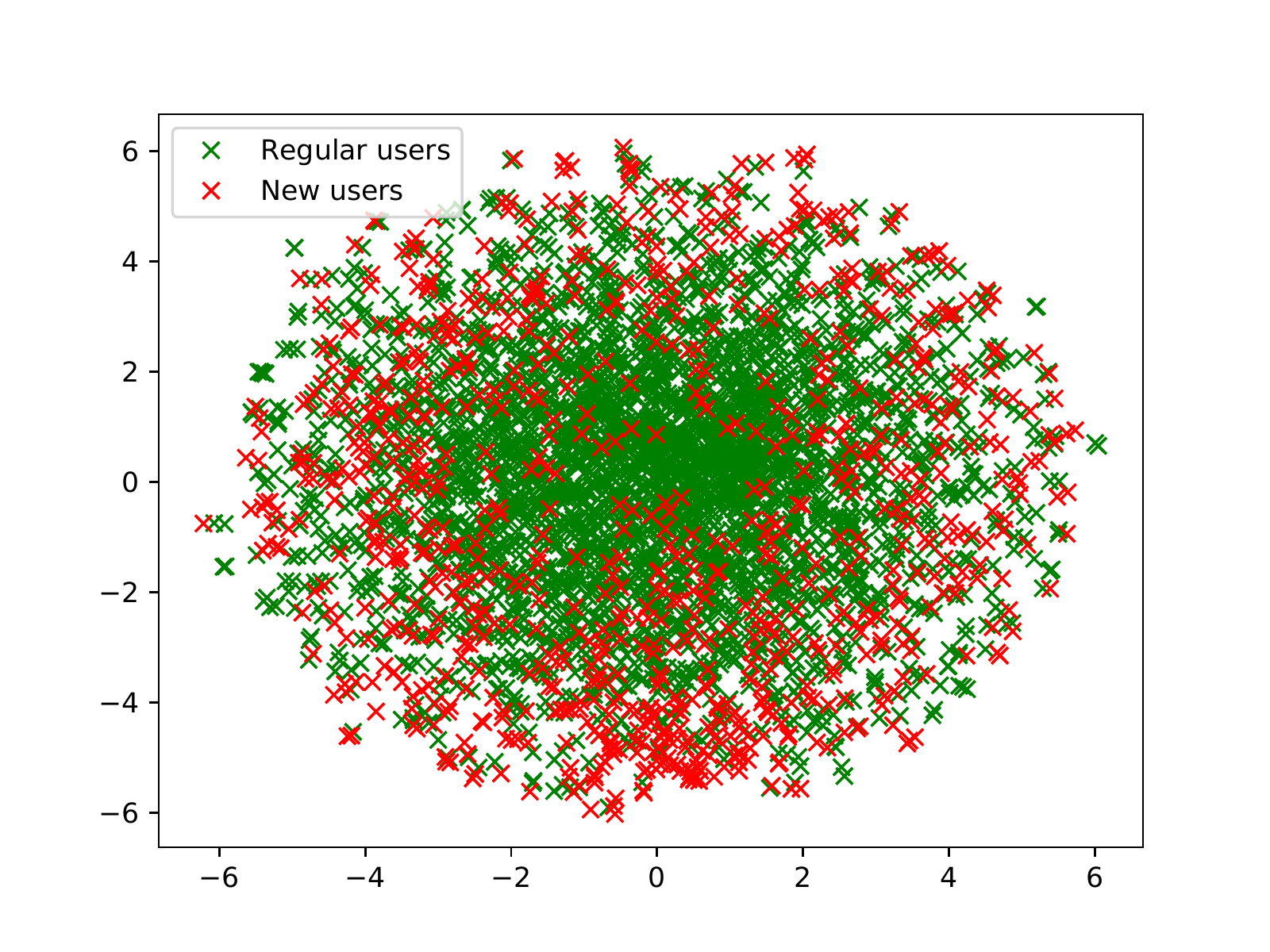}
}\hspace{-13mm}
\quad
\subfigure[w/o Meta Learner]{
\includegraphics[width=7.2cm]{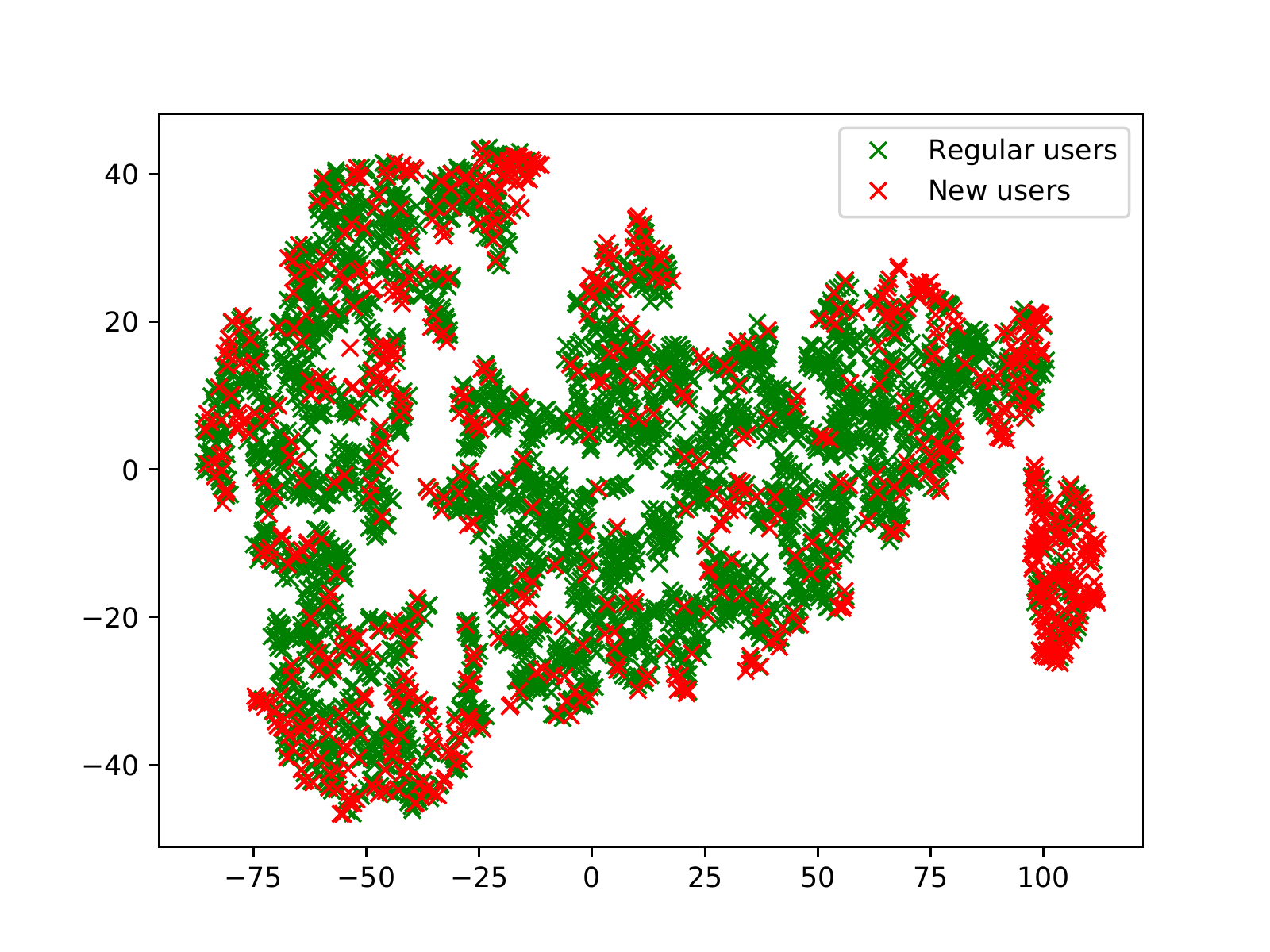}
}
\caption{Visualization of new/regular users’ preference embedding \textbf{WITH fine-tuning}.}
\label{fig:fine-tuning}
\end{figure*}

\textbf{More consistent embedding space.}

Furthermore, as shown in Fig.\ref{fig:w/o fine-tuning}-\ref{fig:fine-tuning}, we visualize the new/regular users preference embedding (outputs of the model) in different situations via t-SNE, we have the following main findings:
\begin{enumerate}
    \item Overall, the users preference embedding in metaCSR modal are both more evenly and regularly distributed than ``w/o Meta Learner".
    
    \item Fine-tune is to further train on a pre-trained model to improve the generalization and performance of the model on the specific new task. We visualize the user preference embedding of regular users and new users without fine-tuning (Fig.~\ref{fig:w/o fine-tuning}) and with fine-tuning (Fig.~\ref{fig:fine-tuning}). The red ``x'' mark represents new users. We can observe that 
    % the distribution of new users' preference of metaCSR (c/d) is contiguously more uniform than ``w/o Meta Learner" (g/h). Also, 
    the distribution of new users' preference of metaCSR changes very little from ``without fine-tuning'' to ``with fine-tuning'', while the distribution through ``w/o Meta Learner'' model changes a lot. This observation shows that the metsCSR algorithm does not need to fine-tune to achieve good performance, that is to say, the pre-trained model has good generalization.
    
    \item Distribution of new users' and regular users' preference embedding have a high degree of mergence in metaCSR (Fig.~\ref{fig:w/o fine-tuning}-\ref{fig:fine-tuning} (a)). It demonstrates that the prior knowledge the model learned from regular users is effectively migrated to the new users. Especially in the ``without fine-tuning'' case in metaCSR, the outstanding performance fully illustrates the high generalization of the model. The analysis above proves that our metaCSR model has better generalization ability. 
\end{enumerate}

% \begin{figure*}[!htbp]
%     \centering
%     \includegraphics[width=1\textwidth]{figs/visualization.pdf}
%     \caption{Visualization of users preference embedding and items embedding in different situations of MovieLens-1M dataset.}
%     \label{fig:visualization}
% \end{figure*} 

% %--------------eight subgraph---------------
% \begin{figure*}[htbp]
% \centering
% \subfigure[pic1.]{
% \includegraphics[width=6cm]{figs/[1]meta_finetune_new.pdf}
% }
% \quad
% \subfigure[pic2.]{
% \includegraphics[width=6cm]{figs/[1]meta_finetune_old.pdf}
% }
% \quad
% \subfigure[pic3.]{
% \includegraphics[width=6cm]{figs/[1]nometa_finetune_new.pdf}
% }
% \quad
% \subfigure[pic4.]{
% \includegraphics[width=6cm]{figs/[1]nometa_finetune_old.pdf}
% }
% \subfigure[pic5.]{
% \includegraphics[width=6cm]{figs/[2]meta_finetune_oldnew.pdf}
% }
% \quad
% \subfigure[pic6.]{
% \includegraphics[width=6cm]{figs/[2]meta_nofinetune_oldnew.pdf}
% }
% \quad
% \subfigure[pic7.]{
% \includegraphics[width=6cm]{figs/[2]nometa_finetune_oldnew.pdf}
% }
% \quad
% \subfigure[pic8.]{
% \includegraphics[width=6cm]{figs/[2]nometa_nofinetune_oldnew.pdf}
% }
% \caption{Visualization of users preference embedding and items embedding in different situations of MovieLens-1M dataset.}
% \label{fig:visualization}
% \end{figure*}

\subsection{Parameter Analysis}

\begin{figure*}[htbp]
\centering
\subfigure[In cold-start scenario.]{
\includegraphics[width=6.5cm]{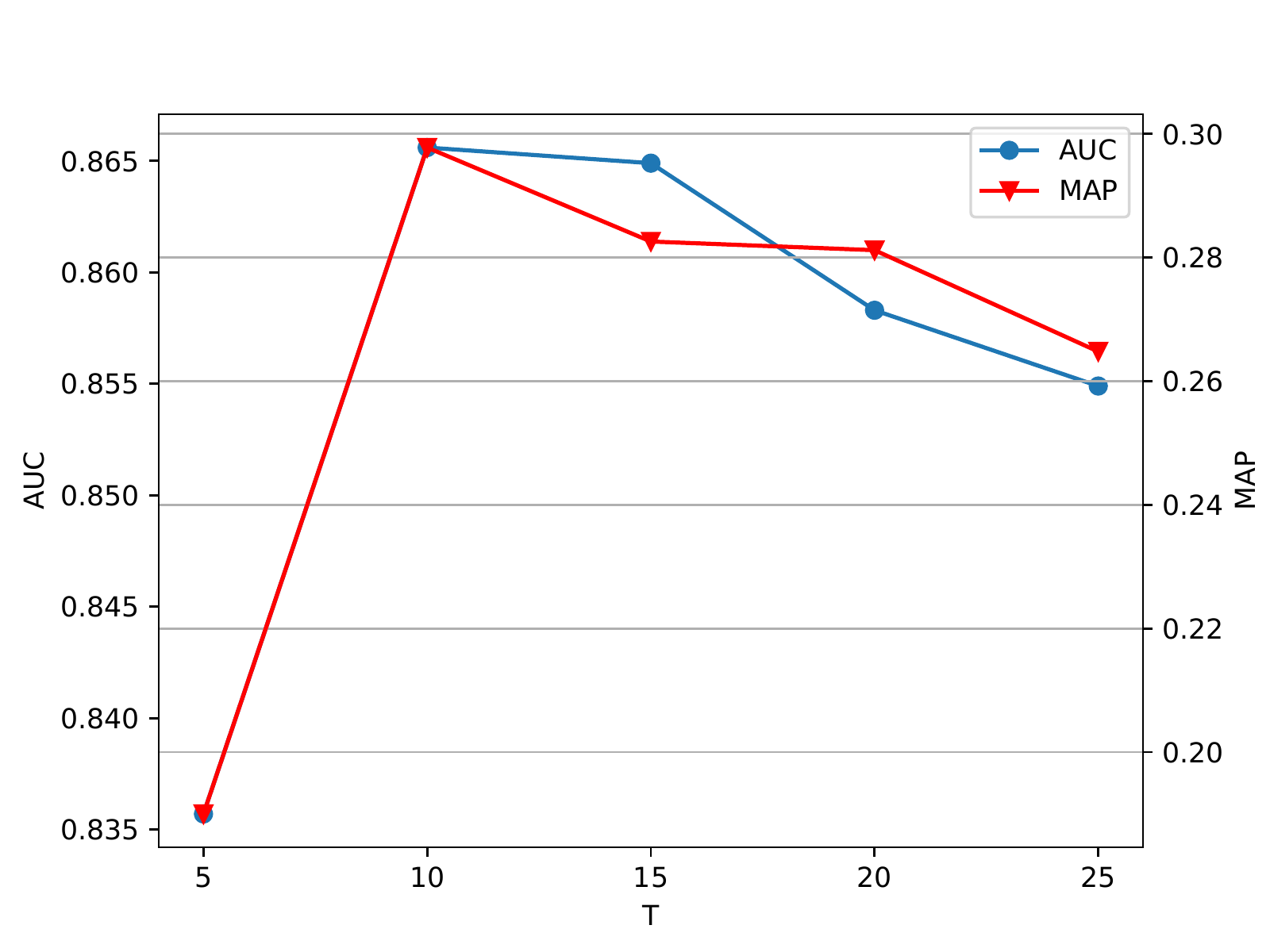}
}\hspace{0mm}
\quad
\subfigure[In warm-start scenario.]{
\includegraphics[width=6.5cm]{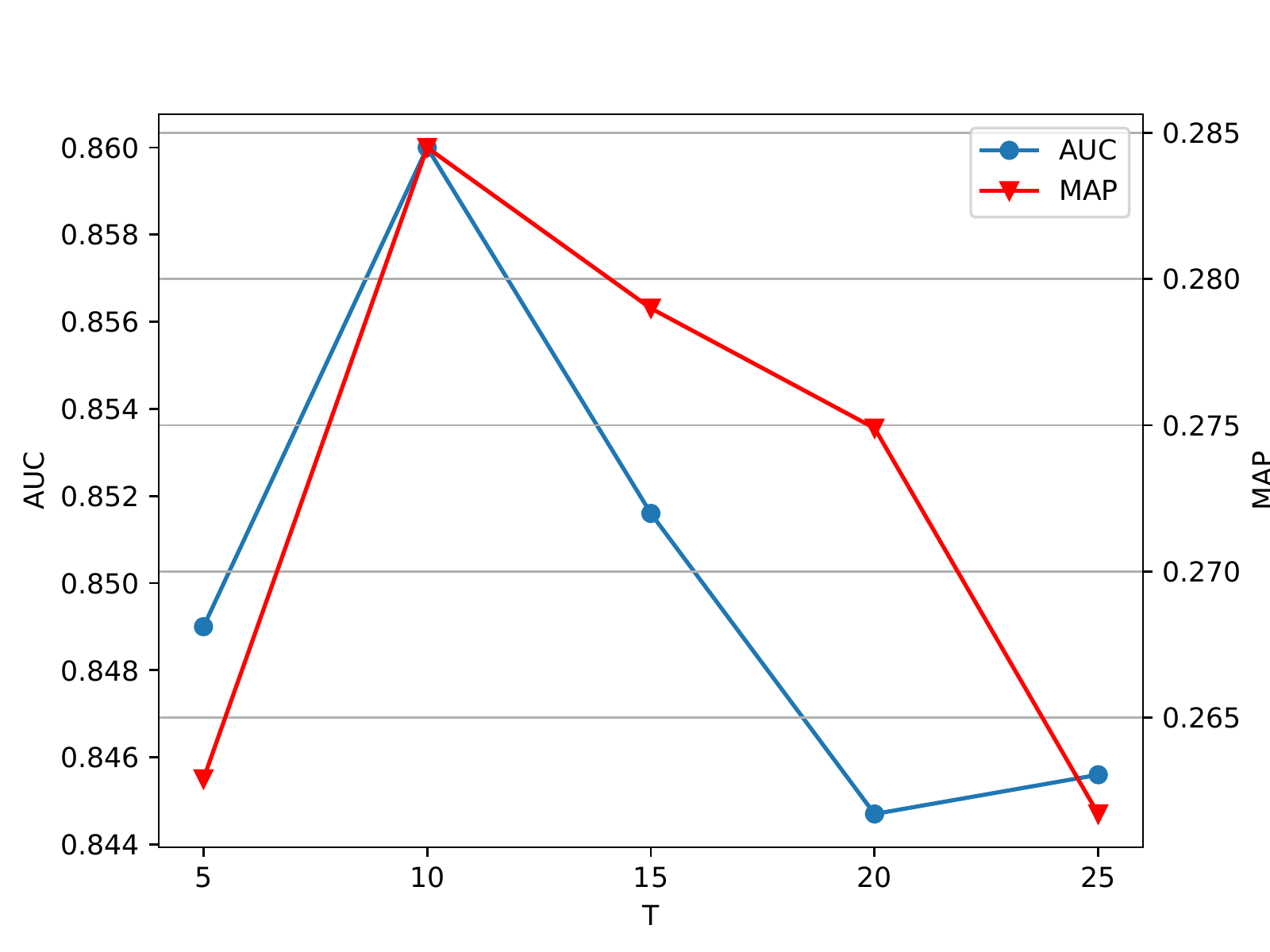}
}
\caption{The performances of different input sequence length $T$ across the evaluation metrics AUC and MAP. It is validated on MovieLens-1M dataset in cold-start and warm-start scenarios. The best performance happened in T=10.}
\label{fig:parameters_analysis}
\end{figure*}

In terms of common sense, the next item that the user will interact with is affected more by the recent behaviors than by the older behaviors. Many behaviors with long intervals may introduce noise information into the system. Thus, our hypothesis is that the appropriate length of the user’s historical records as the input of the sequential recommender is helpful to accurately predict the user’s dynamic preference. To empirically investigate the impact of the length $T$ of behavior sequences on recommendation performance, we conducted the parameter analysis with $T$ in {5, 10, 15, 20, 25}, the results are shown in Fig. \ref{fig:parameters_analysis}. It demonstrates the rationality of our hypothesis, and the appropriate input sequence length is 10. 
Also, since we regard the cold-start recommendation task as the few-shot learning task, choosing a shorter sequence length is more in line with the experimental setting of few-shot learning.

\section{Conclusion and Discussion}

In this paper, we propose a meta-learning based cold-start sequential recommendation framework called metaCSR, which can learn the common patterns from regular users' behaviors and optimize the initialization so that the model can quickly adapt to new users after one or a few gradient updates to achieve optimal performance. The extensive experiments on three widely used datasets show the remarkable performance of metaCSR in dealing with user-cold start problem. Meanwhile, in addition to quantitative analysis, we also qualitatively analyze the model efficacy. The analysis results show that our model has good generalization.

Our proposed metaCSR is a general framework for cold-start sequential recommendations. It does not require any additional side information other than user ID, item ID, and user-item interaction matrix. Certainly, additional auxiliary information, such as item attributes, user profiles, and even context information, multi-modal features, can be seamlessly integrated into our framework through entity representation or sequence modeling to further improve model performance. For efficiency reasons, we use the self-attention-based model as the sequential recommender in this paper, but metaCSR has high applicability since our diffusion representer and meta learner are universal modules, which can be adapted into any sequential recommendation model, even in traditional recommendation methods. The meta-learning way is naturally suitable for cold-start recommendation issues. In the future, we will focus more on the optimization of meta-learner to further improve optimization efficiency and model generalization.

\begin{acks}
This work is supported by the National Key R\&D Program of China (2018AAA0100604), the Fundamental Research Funds for the Central Universities (2021RC217), the Beijing Natural Science Foundation (JQ20023), the National Natural Science Foundation of China (61632002, 61832004, 62036012, 61720106006).
\end{acks}

\bibliographystyle{ACM-Reference-Format}
\bibliography{metaCSR}

\end{document}